\documentclass[fleqn,usenatbib]{mnras}
\include{notations}
\usepackage{slashed}
\usepackage{graphicx}
\usepackage{grffile}
\usepackage[normalem]{ulem}
\usepackage{dcolumn}
\usepackage{xcolor}
\usepackage{booktabs}
\usepackage{amsmath}
\usepackage{amssymb}
\usepackage{slashed}
\usepackage{todonotes}
\usepackage{multirow}
\usepackage{hyperref}

\usepackage[T1]{fontenc}
\usepackage{ae,aecompl}

\usepackage{graphicx}	
\usepackage{amsmath}	
\usepackage{amssymb}	

\title[Explaining Cosmological Anisotropy]{Explaining Cosmological Anisotropy: \\ Evidence for Causal Horizons from CMB data}

\author[Fosalba \& Gazta\~{n}aga]{
Pablo Fosalba$^{1,2}$\thanks{e-mail:fosalba@ice.csic.es}
and Enrique Gazta\~{n}aga,$^{1,2}$
\\
$^{1}$ Institute of Space Sciences (ICE, CSIC), Campus UAB, Carrer de Can Magrans, s/n, 08193 Barcelona, Spain\\
$^{2}$ Institut d'Estudis Espacials de Catalunya (IEEC), Carrer Gran Capit\`a 2-4, 08193 Barcelona, Spain
}

\date{Accepted XXX. Received YYY; in original form ZZZ}

\pubyear{2020}

\begin{document}
\label{firstpage}
\pagerange{\pageref{firstpage}--\pageref{lastpage}}
\maketitle

\begin{abstract}
The origin of power asymmetry and other measures of statistical anisotropy on the largest scales of the universe, as manifested in Cosmic Microwave Background (CMB) and large-scale structure data, is a long-standing open question in cosmology. In this paper we analyze the Planck Legacy temperature anisotropy data and find strong evidence for a violation of the Cosmological principle of isotropy, with a probability of being a statistical fluctuation of order $\sim 10^{-9}$. The detected anisotropy is related to large-scale directional $\Lambda$CDM cosmological parameter variations across the CMB sky, that are sourced by three distinct patches in the maps with circularly-averaged sizes between $40$ to $70$ degrees in radius. We discuss the robustness of our findings to different foreground separation methods and analysis choices, and find consistent results from WMAP data when limiting the analysis to the same scales. We argue that these well-defined regions within the cosmological parameter maps may reflect finite and casually disjoint horizons across the observable universe.  In particular we show that the observed relation between horizon size and mean dark energy density within a given horizon is in good agreement with expectations from a recently proposed model of the universe that explains cosmic acceleration and cosmological parameter tensions between the high and low redshift universe from the existence of casual horizons within our universe.

\end{abstract}





\section{Introduction}
\label{sec:intro}

The standard cosmological model stands on the shoulders of a fundamental assumption: that the universe is statistically homogeneous and isotropic on the largest scales. This assumption has been thoroughly tested over the last years both with Cosmic Microwave Background (CMB) and Large-scale structure data. In particular, the analysis of CMB data, most notably from the WMAP \citep{WMAP9} and Planck \citep{Plegacy} experiments, has not yet provided conclusive evidence for the hypothesis of Cosmological Isotropy (\citealt{H04,2005ApJ...618L..63H,E07,2007MNRAS.378..153L,H09,2009MNRAS.396..511S}; see also \citealt{P18isotropy} and references therein). Moreover, Galactic foreground contamination or known systematic effects in the data alone can not explain the observed CMB "anomalies", i.e, large-scale deviations from the concordance $\Lambda$CDM model (see e.g, \citealt{2014JCAP...08..006R}; see \citealt{P18isotropy} for a recent overview). Power asymmetry from CMB data has also been a matter of intense debate and scrutiny (\citealt{ 1998MNRAS.295L..35G,E07,2008JCAP...09..023L,2009ApJ...699..985H,2010MNRAS.407..399P,2013ApJ...773L...3A,2019JCAP...08..007S}, see also \citealt{2013PhRvD..87l3005D} for a comprehensive discussion and references therein), and evidence has been reported that this could source deviations from isotropy on cosmological scales \citep{H09}. However, a more recent analysis based on Planck data finds no evidence for such power asymmetry when all scales are taken into account \citep{2015JCAP...01..008Q}. This is in qualitative agreement with the latest results from the Planck Collaboration analysis \citep{P18isotropy} where they conclude that the observed power asymmetry is not robust to foreground contamination or systematic residuals.
It is important to note that previous analysis have concentrated on quantifying potential deviations from statistical isotropy using a statistical prior. First analyses using WMAP data looked for the direction of maximal asymmetry in the sky, thus quantifying anisotropy for a given preferred direction \citep{H09}. In turn this led to proposing a particular angular distribution of power in the sky to simply capture the observed anisotropy, such as the so-called "dipole anisotropy" modulation (\citealt{2005PhRvD..71h3508P,2007ApJ...656..636G}). This same model has been further constrained with Planck data (\citealt{P13isotropy,P15isotropy,2015PhRvD..92f3008A,2016JCAP...06..042M,P18isotropy}). Alternatively, a recent analysis \citep{2018arXiv181208980H} focuses on quantifying possible CMB peak shifts across the sky, finding significant variations, but they attribute this behaviour to possible systematic effects or the solar dipole. Complementary evidence for cosmological anisotropy has been investigated using probes of the low redshift universe (see \citealt{2011MNRAS.414..264C, 2020arXiv200914826S} and references therein).

In this paper we re-assess Cosmological Isotropy using the latest Planck and WMAP datasets from a different angle, as we do not use any prior in our approach. In particular, we focus on looking at possible variations of the best-fit cosmological parameters across the sky, and test the robustness of our findings to our analysis choices, data-cuts and foreground contamination. As we shall discuss below, our analysis shows compelling statistical evidence for large-scale anisotropies (Gaussian isotropic hypothesis has a probability of $\sim 10^{-9}$), sourced by large-scale directional variations in all the basic $\Lambda$CDM parameters. We argue that a possible source for such anisotropies is the existence of primordial causal horizons within our observable universe. Such horizons can result from inflation and explain the observed late time cosmic acceleration  \citep{G20,Gazta2021}. As discussed below, the measured correlation between horizon size and mean dark-energy density within the detected horizons turns out to be in good agreement with expectations from this model.

Previous analyses looking for directional dependence of the cosmological parameters involved WMAP \citep{2013ApJ...773L...3A} and, more recently, Planck data \citep{2018JCAP...01..042M}.  \cite{2013ApJ...773L...3A} focused on the power asymmetry and thus, their approach is different with respect to ours (i.e, theirs is prior dependent). They did find a $3.4 \sigma$ evidence of power asymmetry but only a hint of directional parameter dependence in some of the basic $\Lambda$CDM parameters, although this non-detection could be due to the lower signal to noise of the WMAP data with respect to Planck and, in particular, the limited range of scales used, $\ell < 600$. On the other hand, \cite{2018JCAP...01..042M} follow a closer methodology to ours, but they use an approximation (first order Taylor expansion) to relate power spectra to the underlying cosmological parameters. Besides, they divide the footprint in a very limited number of patches to sample the sky and do not include any residual foreground parameter that affect the high multipoles ($\ell > 900$) of the power spectra, as we do, what can explain that they do not find conclusive evidence for a directional parameter dependence from their analysis. In fact, they only detect some significant anisotropic signal for those patches of the sky that have very small area after removing the overlap with the Galactic mask, which is at variance with what we find, as we shall discuss in detail below.

The paper is organized as follows: Section \S\ref{sec:method} describes our data analysis pipeline, its validation and the analysis choices taken. In Section \S\ref{sec:results} we present our main results, including the evidence for spatially coherent cosmological parameter variations across the sky, and the associated evidence for what we dub as "horizons" (see text for details) and analyze its robustness against foreground contamination.  
We conclude in Section \S\ref{sec:discussion} with a summary of our main results, a discussion of their implications and point out possible future directions.

\section{Methodology and Data analysis pipeline}
\label{sec:method}

\subsection{Analysis Pipeline}
\begin{figure}
 \includegraphics[width=0.48\textwidth]{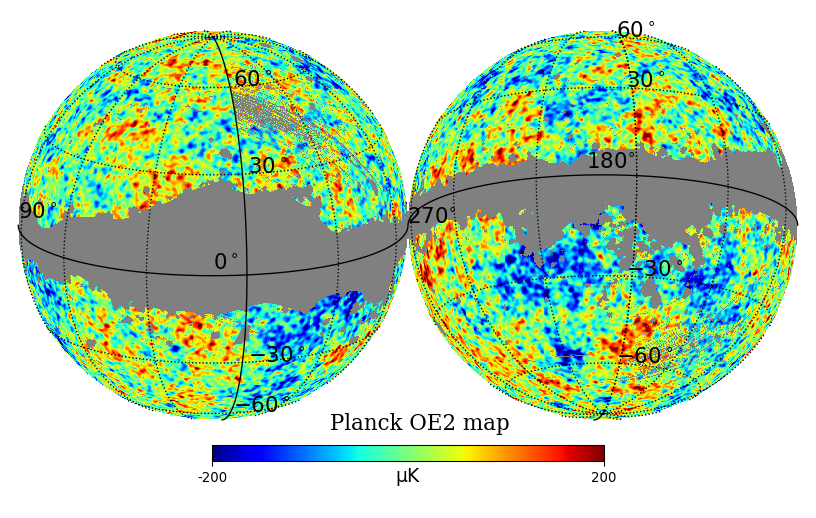}
 \caption{Planck 2018 half-mission Odd-Even (OE) SMICA temperature anisotropy map, smoothed with 1 degree FWHM Gaussian beam, shown in orthographic projection. In this projection, the left hemisphere is centered at the Galactic center, whereras the right hemisphere is around the anti-Galactic center. The Galactic mask used is overlaid (in grey color).}
  \label{fig:planck_sm1deg}
\end{figure}

In order to analyze the Planck temperature maps and investigate possible variations of the best-fit cosmological parameters as a function of position in the sky, we proceed as follows:

\begin{itemize}
    \item {\it Step 1: Data}. For our main analysis, we make use of the Planck 2018 "Odd-Even" ring half-mission temperature anisotropy maps (OE maps)  obtained with the SMICA foreground separation method, along with the Galactic mask given by the so-called "common mask" in combination with the half-mission Odd-Even missing pixels (i.e, missing rings of data), which leaves about $76\%$ of the sky available for the cosmological analysis. All the data used in the analysis has been downloaded from the Planck Legacy Archive\footnote{\tt{https://pla.esac.esa.int/}}. Fig.\ref{fig:planck_sm1deg} shows one of the two half-mission Odd-Even (OE) temperature anisotropy maps used in our main analysis, smoothed with a 1 deg. FWHM Gaussian beam for better visualization, with the Galactic mask overlaid
    \item {\it Step 2: Survey sub-samples}. We set the size of the disc where we perform the cosmological analysis. We choose to use a circular shape as it is a simple symmetric mask which also minimizes mask de-convolution effects on the resulting map angular power spectrum. For convenience, the location of the disc centers are chosen as the coordinates of the pixel centers in a (low-resolution) HEALPixix map. We then project a given disc onto the Planck footprint, defining the survey patch as those pixels of the disc which do not overlap with the Planck galactic foreground emission and resolved point source mask 
    \item {\it Step 3: Power spectrum estimation}. We measure the angular power spectrum $C_\ell$'s of the Planck temperature map in the survey patch (i.e disc). We use the multipole range from $\rm{\ell_{min} = 32}$ to $\rm{\ell_{max}=2000}$, to minimize foreground residuals present in Planck data at low and high multipoles, and make the Gaussian assumption an accurate approximation to the true likelihood (for $\ell > \ell_{min}$). Following the official Planck analysis papers, we compute $C_\ell$'s band-powers with $\Delta \ell = 30$, so that off-diagonal elements of the covariance matrix are kept at the $\sim 10 \%$ level for the full Planck footprint (and somewhat larger for smaller patches), but ignoring them do not bias cosmological parameters within 1-$\sigma$ statistical errorbars (see e.g, Table \ref{tab:cosmo})
    \item {\it Step 4: Error estimation}. Diagonal $C_\ell$'s errors are estimated by re-scaling those from the official Planck power spectrum by the effective area of the patch used, i.e, ${\Delta} C^{\cal D}_{\ell}/{\Delta} C^{Planck} = \sqrt{f^{Planck}_{sky}/f^{\cal D}_{sky}}$, with $f^{Planck}_{sky}=0.57$ which is a good estimate of the "mean" effective area from the masks of the HFI frequency channels (100,143 and 217 GHz) included in the Planck likelihood. In our fiducial analysis case (sub-areas of 60 degrees in diameter), this corresponds to a mean fraction of the sky available for the cosmological analysis (i.e, not overlapping with the Galactic mask), $f^{\cal D}_{sky} \simeq 0.05$, although there are variations from disc to disc across the sky, which we take into account 
    \item {\it Step 5: Cosmological parameter estimation}. Finally, we find the best-fit base $\Lambda$CDM cosmological parameters (plus one nuisance foreground residual parameter, see below for details) to the measured $C_\ell$'s and errors for each disc. We repeat this operation for all discs that sample the full Planck footprint
\end{itemize}

In Section \S\ref{sec:choices} below, we describe in detail the particular choices we make for the main analysis as well as the codes used for the power spectrum and likelihood estimation.

\subsection{Analysis Choices}
\label{sec:choices}

In what follows we shall use the SMICA CMB map, as it is the foreground cleaning method of choice (ie, reference CMB map) in the Planck cosmology papers (\citealt{P18cosmo,Plegacy}). However, we shall show below that our results are robust to the choice of foreground separation algorithm used. As for the data cuts, we choose to use the "odd-even" ring half mission maps (OE maps), where frequency maps are built using every other pointing period or ring, restricting to either
the odd- or the even-numbered rings. The OE maps thus have noise which is largely uncorrelated and free of the scanning-dependent systematics. This is important to avoid possible biases in the inferred cosmological parameters from the small scales in the maps, ie. high multipoles in the $C_\ell$ (Karim Benabed, private communication; see also Figure A2 in appendix of \citealt{2020A&A...641A...5P}). 

The fiducial analysis value of the disc size is set to 60 degrees in diameter, which yields on average $\sim$ 2100 sq.deg. of CMB sky outside the galactic and point-source mask. This mask size turns out to be optimal in terms of signal-to-noise, as we will argue below, but we have checked that our main results are robust to changes in disc area by a factor of 5 (ie. disc sizes from 40 to 90 degrees in diameter). As for the resolution of the map pixels used as centers of the set of disc-masks, in our main analysis we choose to work at HEALPixix resolution $\rm{Nside=4}$ (i.e, 192 pixels across the sky) since this already captures all the significant variations of cosmological parameters across the sky (ie. $\ell < 12$), as it will be shown. However, when estimating the size of such "horizons" in the parameter maps we shall use finer resolutions ($\rm{Nside=16}$ and $32$, what results in 3072 and 12888 discs across the sky, respectively), to have better statistics (see \S\ref{sec:horizonsize}). 

The angular power spectrum in each disc is computed using the $\tt{PolSpice}$ code \citep{Spice,PolSpice} which has been extensively used in previous CMB analysis (see e.g, \citealt{2004ApJ...617L..95F,2018ApJ...869...38H}). $\tt{PolSpice}$ is a an approximate pseudo-Cl method that is unbiased, and allows for the fast and accurate estimation of the 2-point correlation functions of pixelated maps, correcting for complicated angular masks of finite sky experiments. 

Finally, best-fit cosmological parameters are inferred from the measured $C_\ell$'s in each disc with $\tt{iMinuit}$ \footnote{https://iminuit.readthedocs.io/en/latest/}, a code for maximum-likelihood fits of statistical models to data that also provides model parameter errors from likelihood profile estimation. $\tt{iMinuit}$ assumes a Gaussian distribution of the model parameters, which is a very good approximation in our case, given the analysis choices we make (we use a sufficiently large low multipole cut in the analysis, $\rm{\ell_{min}=32}$, see \S\ref{sec:method}). $\tt{iMinuit}$ has been widely used for CMB in general and for Planck data analysis in particular in recent years \citep{CAMEL,2016A&A...594A..11P}, where its validation with respect to traditional MCMC methods has been thoroughly discussed \footnote{see the CAMEL project wiki page for details: http://camel.in2p3.fr/wiki/pmwiki.php}.  The advantage of $\tt{iMinuit}$ over MCMC methods lies in its speed, typically orders of magnitude faster than traditional Bayesian methods but with comparable accuracy, provided the parameters are Gaussian distributed. The gain in speed is key in our analysis since a full MCMC approach would be prohibitive given the large number of pixels for which we have to get cosmological parameter fits.

As a proof of method, in this paper we concentrate on the basic flat-space $\rm{\Lambda CDM}$ cosmological parameters ($\Omega_c\rm{h^2}, \Omega_b\rm{h^2}, \rm{H_0}, \rm{n_s}, \rm{A_s}$), and setting the optical depth to reionization, $\tau$, fixed. We note that this choice does not limit the generality of our results in practice, since when analyzing the temperature data alone, the optical depth is tightly correlated to the primordial amplitude parameter, $\rm{A_S}$, which we do leave free in our analysis. We fix all parameters that are not sampled to the following values: $\rm{\tau = 0.0522}$, $\rm{\sum m_{\nu} = 0.06 \, eV}$, $\rm{N_{eff}= 3.04}$ and $\rm{r = 0}$ (i.e, no tensors).
Additionally, we include an "effective foreground residual" parameter ($\rm{A^{PS}_{eff}}$), that accounts for the combined contamination from the Cosmic Infrared Background (CIB) and un-resolved extragalactic point sources \citep{2020A&A...641A...5P}, the two main sources of contamination in Planck HFI data at large multipoles (ie., small angular scales) \footnote{We note the "abuse" of language in our notation as our nuisance parameter, $\rm{A^{PS}_{eff}}$,  effectively combines, in the Planck data release papers notation, the following set of nuisance parameters: the extragalactic Poisson point source contribution from the different HFI frequency channel pairs $\rm{100,143,217}$ GHz ($\rm{A^{PS}_{100}}, \rm{A^{PS}_{143}}, \rm{A^{PS}_{143x217}}, \rm{A^{PS}_{217}}$), and the CIB amplitude at $\rm{217}$ GHz ($\rm{A^{CIB}_{217}}$), as described in \citealt{2020A&A...641A...5P}, Table 16}. These combined residual foregrounds effectively behave as a single "shot-noise" contribution, $\rm{A^{PS}_{eff}}$  at the power spectrum level if one limits the analysis to $\ell < 2000$ \citep{2016A&A...594A..11P}. In what follows, we shall denote this nuisance parameter simply as 
$\rm{A_{PS}}$ for the ease of notation.

As for the multipole range used, we do not consider the lowest multipoles and set $\rm{\ell_{min}=32}$. This makes our analysis closest to the so-called Planck high-$\ell$ likelihood \citep{2020A&A...641A...5P}, and it has a two-fold advantage: first it removes scales which are in principle more sensitive to residual foreground contamination from an imperfect modeling of diffuse galactic emission (e.g, dust) and secondly, it makes the assumed Gaussian likelihood of the power spectrum a good approximation. At the high multipole end, we cut at $\rm{\ell_{max}=2000}$, since this makes the treatment of foreground residuals simpler in terms of effectively one single parameter, as mentioned above, without losing any significant cosmological constraining power with respect to the full Planck data resolution, ie. $\rm{\ell_{max}=2500}$. We note that in cutting at $\ell_{max} =2000$ we can also safely neglect the {\it clustered} CIB component which deviates from the purely shot-noise behaviour (it scales as a power law) assumed in our analysis (see Section \S 3.3.1 in \citealt{2020A&A...641A...5P}, see also Figure \ref{fig:clscompmap} below).

\subsection{Pipeline Validation}
\label{sec:pipevalid}

We next validate our pipeline by comparing results using our power spectrum (with $\tt{PolSpice}$) and likelihood estimation ($\tt{iMinuit}$) agree with the official results published by the Planck collaboration for the fiducial $\Lambda$CDM  parameters, when restricted to temperature data (and low-multipole polarization, ie., TT+LowE in the notation used in the Planck data release). In addition we provide an estimate of the corresponding high-multipole residual foreground contamination amplitude, $\rm{A_{PS}}$, which depends on the particular $\rm{\ell_{max}=2000}$ choice we use, and therefore not directly comparable to the Planck official results.

\begin{figure}
 \includegraphics[width=0.48\textwidth]{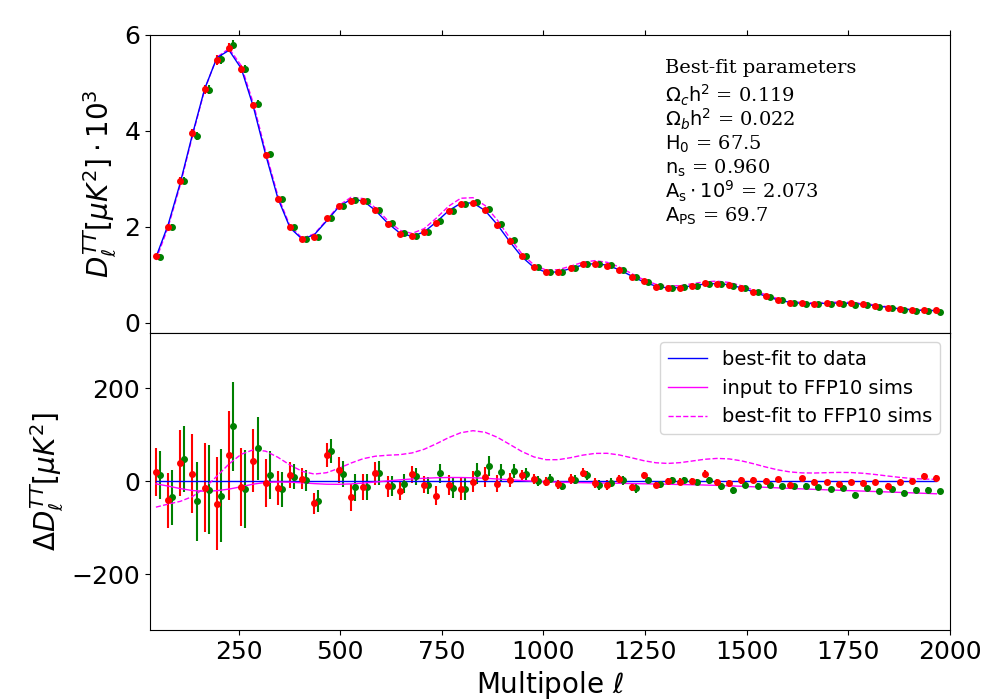}
 \caption{Upper panel: angular power spectrum of the Planck SMICA CMB temperature map (red symbols with errors), and its corresponding best-fit to a $\Lambda$CDM cosmology (blue solid line). Published results from the Planck collaboration are also shown (green symbols, slightly shifted to the right for clarity). For reference, we also show the input (solid magenta) and the corresponding best-fit to the FFP10 simulations (dashed magenta; see Section \S\ref{sec:sims} for details). Lower panel: relative deviations with respect to the best-fit model} 
 \label{fig:clfit}
\end{figure}

\begin{figure}
 \includegraphics[width=0.48\textwidth]{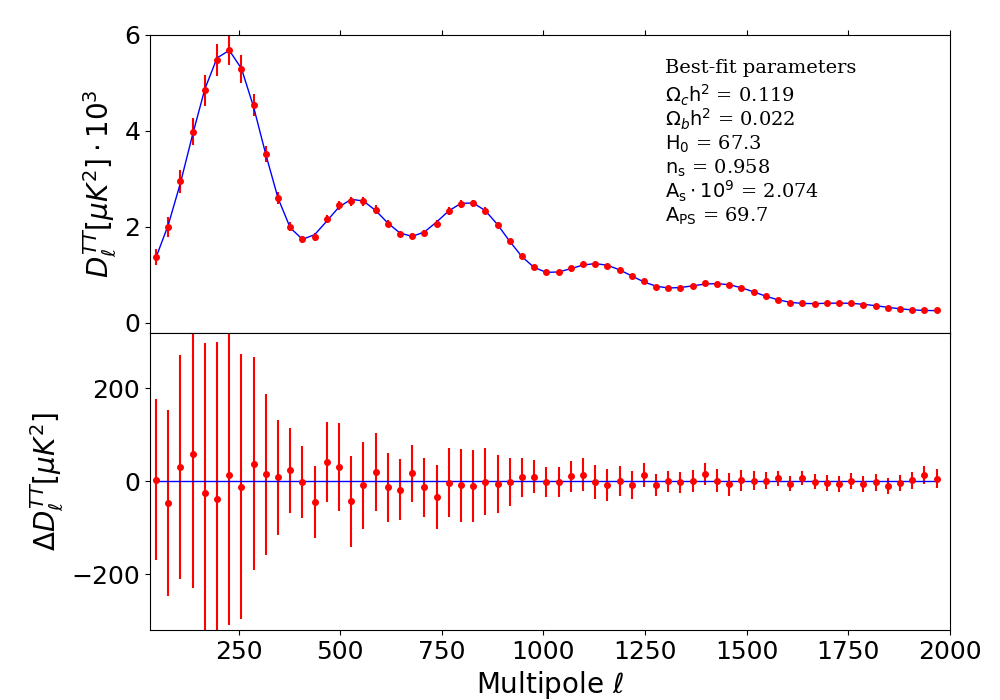}
 \caption{Same as Figure \ref{fig:clfit} for the mean power spectrum over discs of 60 degree diameter across the sky} 
 \label{fig:clfitdisc}
\end{figure}

Figure \ref{fig:clfit} shows the angular power spectrum for the Planck SMICA temperature map as measured with our pipeline (red symbols), compared to the published results by the Planck Collaboration (green symbols). Errorbars for both cases are the ones provided by the Planck Collaboration. The very good agreement between our $C_{\ell}$'s and the official Planck results validates our pipeline. Moreover the slight discrepancy at very high multipoles ($\ell > 1500$) does not impact our results, as we shall discuss in Section \S\ref{sec:cuts}. Figure \ref{fig:clfitdisc} shows the corresponding average $C_{\ell}$'s over the disc sub-areas of 60 degree diameter, $<{\cal D}_{60}>$, used in our main analysis, along with the best-fit $\Lambda$CDM model.

Table \ref{tab:cosmo} summarizes the best-fit cosmological parameters for the different estimates of the power spectrum discussed above. In particular, it shows that our results for the full Planck footprint (2nd column) agree with the best-fit parameters obtained by the Planck collaboration within the 1-$\sigma$ errors (first column; see also Table 2 in \citealt{P18cosmo}). In turn these values are consistent with the fit obtained from the average power spectra, $<{\cal D}_{60}>$,  over the $3072$ discs of 60 degree diameter (see 3rd column), with which we homogeneously sample the sky to assess anisotropies or angular variations in the cosmological parameter fits. Moreover, the fact that the discs we use for the analysis recover cosmological parameters consistent with the full footprint indicates that the cosmological information encoded in the sub-areas is unbiased within the Planck accuracy.

\begin{table}
 \caption{Best-fit cosmological parameters: comparison between the official Planck results for TT+LowE (\citealt{P18cosmo}, see first column of Table 2), results from our pipeline using the full Planck footprint and our analysis choices (see text for details), and same as the latter but for the average over circular sub-areas of 60 degrees in diameter.}
 \label{tab:cosmo}
 \begin{tabular}{lccc}
  \hline
  Parameter & Planck2018 Results & Full footprint &  Discs \\ 
  \hline
  $\Omega_c \rm{h^2}$ & 0.121 $\pm$  0.002 & 0.119 & 0.119 \\
  $\Omega_b \rm{h^2}$ & 0.0221 $\pm$ 0.0002 & 0.022 & 0.022  \\
  $\rm{H_0}$ & 66.9 $\pm$  0.9 & 67.5 & 67.3  \\
  $\rm{n_s}$ & 0.963 $\pm$  0.006 & 0.960 & 0.958 \\
  $\rm{A_s}\cdot 10^9$ & 2.09 $\pm$  0.03 & 2.07 & 2.07  \\
  $\rm{A_{PS}}$ & -  & 69.7  & 69.7   \\
  \hline
 \end{tabular}
\end{table}

\section{Results}
\label{sec:results}

We apply the methodology described in Section \S\ref{sec:method} to the SMICA CMB temperature map, and obtained a set of power spectra for the set of circular patches of a given diameter size, $\cal D$ hereafter, that sample the entire Planck footprint. The set of power spectra obtained from the suite of masks is the input for the cosmological parameter estimation with which we shall assess possible coherent variations of best-fit parameters across the sky.

\subsection{Cosmological parameter maps}
\label{sec:cosmoparams}

\begin{figure*}
 \includegraphics[width=\textwidth]{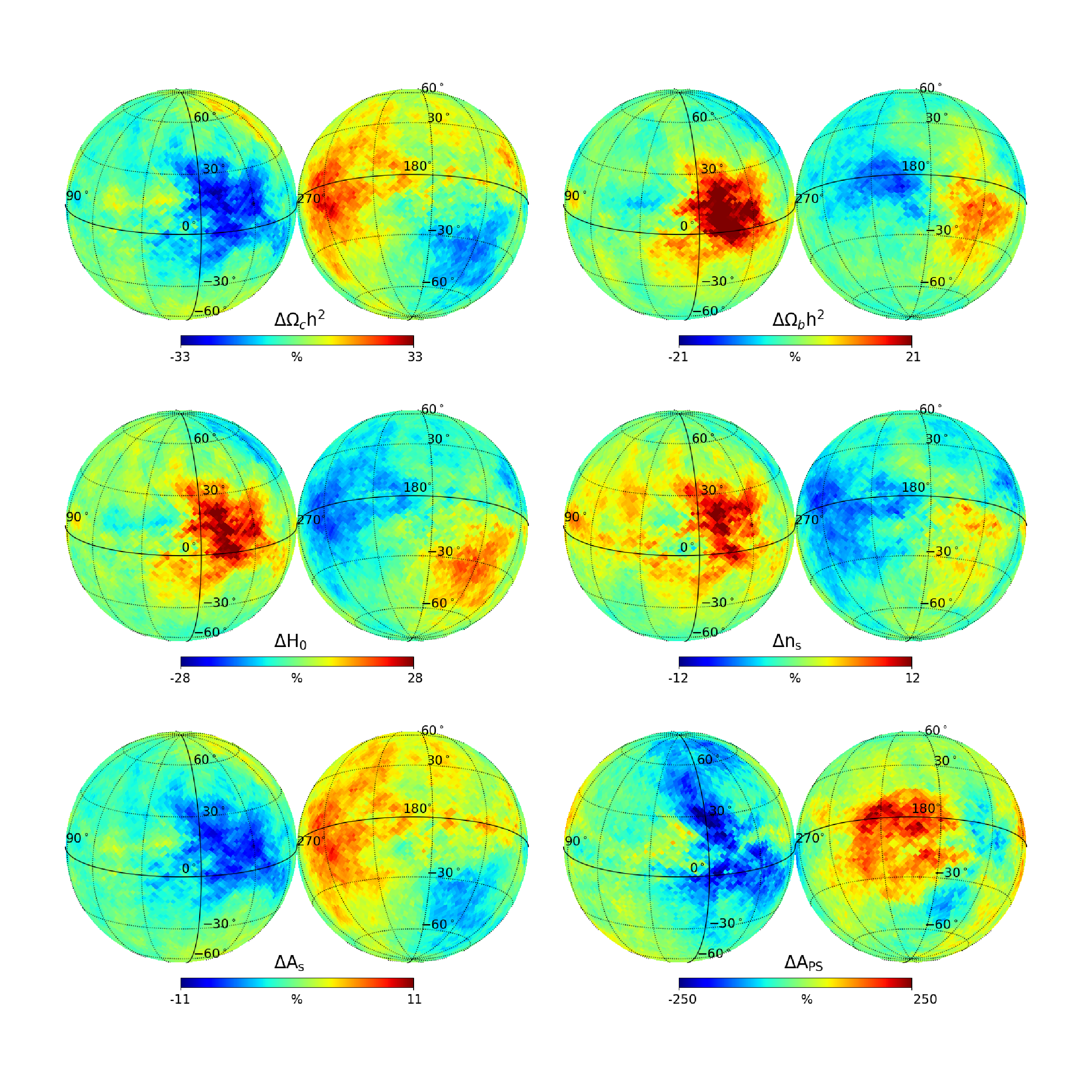}
 \caption{$\Lambda$CDM and residual foreground nuisance parameter ($A_{PS}$) variations across the sky (in Galactic coordinates). In each panel the color-code displays ($\%$) fluctuations with respect to mean over all-sky within the range $\pm 4 \sigma$. The same dynamic range is used for all the parameter maps shown in this paper. This map, which uses $12288$ discs across the sky centered at the pixels of a HEALPix $\rm{Nside} = 32$ resolution map, is the baseline for the main results of the paper, and it is based on the input SMICA temperature map. Results obtained using maps from other foreground separation methods are consistent with those from SMICA, as discussed in Section \S\ref{sec:foregrounds}.} 
 \label{fig:SMICAmaps}
\end{figure*}

Figure~\ref{fig:SMICAmaps} displays the basic $\Lambda$CDM parameter variations across the sky. For visualization purposes, we use a "high-resolution" sampling of the sphere using discs centered at the $12288$ pixels of a HEALPix $Nside=32$ map, for which angular power spectra and cosmological parameters are estimated. Parameter ranges shown are symmetrized and chosen to be illustrative of dynamic range and size of what we describe as "horizons" for the physical implications that we shall describe below. These "horizons" correspond to coherent patches with distinct parameter values above or below the mean over the full sky.

It is clear from these maps that there are common features or angular patterns in them. In particular, all cosmological parameters exhibit similar variations across the sky, although these anisotropic patterns are at least a factor of two larger for the acoustic oscillation related parameters ($\omega_c \rm{h^2}$, $\Omega_b \rm{h^2})$ than those describing the primordial spectrum of fluctuations ($\rm{A_S}, \rm{n_s}$). On the other hand, a rather different (and much larger) spatial pattern of fluctuations is observed for the foreground-residual amplitude, ($\rm{A_{PS}})$, which suggests that the observed "horizons" in the cosmological parameter maps have a different cause than the corresponding features in the foreground parameter map. 

\subsection{Horizon size estimation}
\label{sec:horizonsize}

Once we have found evidence for the existence of "horizons" or coherent patches for the parameter fluctuations around the mean across the sky, we shall estimate the size of such horizons. For this purpose we shall assume that horizon centers are those pixels in the map which maximize the differences with respect to the mean, ie., they are well-defined peaks (global maxima/minima) in the parameter fluctuation map. Therefore we start by finding the peaks in the maps and ranked them according to their height. The largest three peak-heights thus define the center of the corresponding horizons that we found to be statistically significant. The location of these peaks or center of the horizons are given in Table \ref{tab:hlocation} \footnote{In order to determine the horizon center or peak locations in the parameter maps we first smooth the maps with a $10^{\circ}$ FHWM Gaussian beam to remove small-scale noise. This smoothing scale is chosen as modes with $\theta < 10$ deg do not contribute significantly to the parameter map, as we shall show in Section \S\ref{sec:hsignificance}}.

\begin{table}
 \caption{Location of Horizons in Planck Temperature map (in Galactic coordinates)}
 \label{tab:hlocation}
 \begin{tabular}{lcc}
  \hline
  Horizon & Longitude (deg) & Latitude (deg)  \\ 
  \hline
 $\rm{H_1}$ & $345$ & $15$ \\
 $\rm{H_2}$ & $240$ & $-5$ \\
 $\rm{H_3}$ & $150$ & $-40$ \\
  \hline
 \end{tabular}
\end{table}

Our method to estimate the size of such horizons is as follows:
\begin{itemize}
    \item first we draw rings of width $\delta \theta$ defined by the common pixels of two concentric discs of radius 
    $\theta$ and $\theta + \delta \theta$ respectively, around the horizon locations given in Table \ref{tab:hlocation} 
    \item we vary the horizon scale $\theta$ until we find a minimum of the "peak height variation" estimator:
    \begin{equation}
        \Delta P = \{<d P>_{\delta \theta} - <d P>_{all-sky}\}/\sigma_{P_{\delta \theta}} \, ,
          \label{eq:horizon}
    \end{equation}
    where $\rm{<d P>_{\delta \theta}}$ is the mean value of the parameter values within the ring pixels, $\rm{<d P>_{all_sky}}$ is the corresponding mean over the entire sky, and $\rm{\sigma_{P_{\delta \theta}}}$ is the rms scatter over the ring pixel values. 
  
\end{itemize}

\begin{figure*}
 \includegraphics[width=\textwidth]{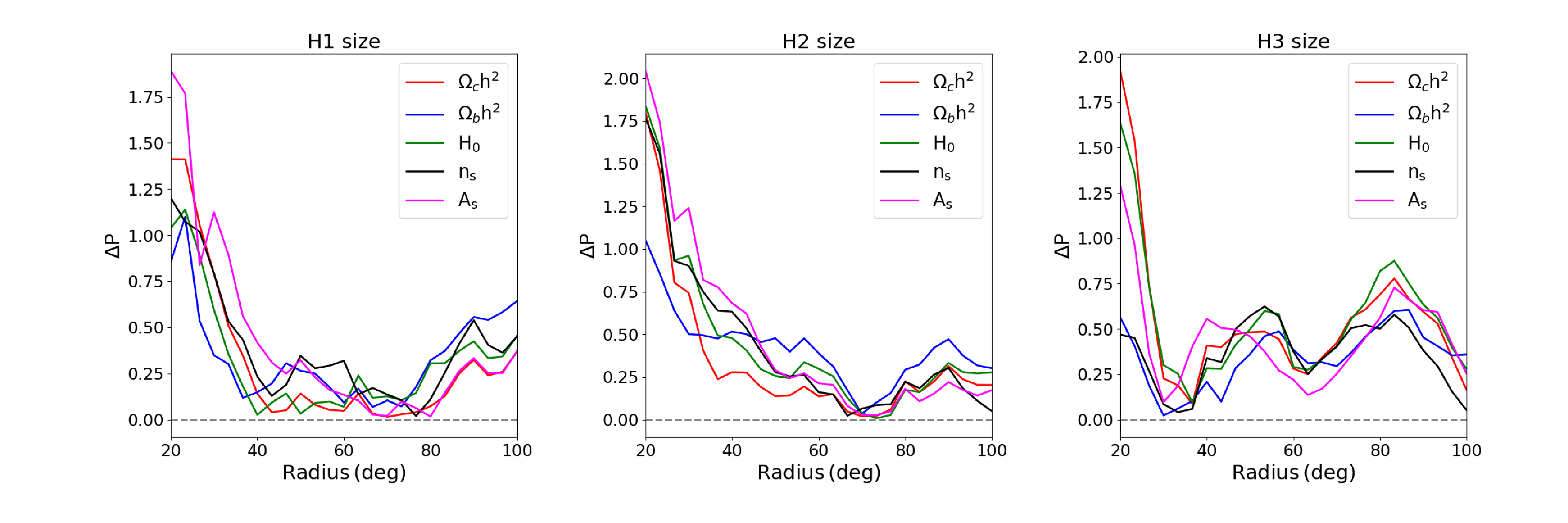}
 \caption{Estimated size of the cosmological Horizons for the 5 $\Lambda$CDM parameters used, when the input power spectra are computed in discs of $40$ degrees in diameter.}
 \label{fig:hsize20}
\end{figure*}

\begin{figure*}
 \includegraphics[width=\textwidth]{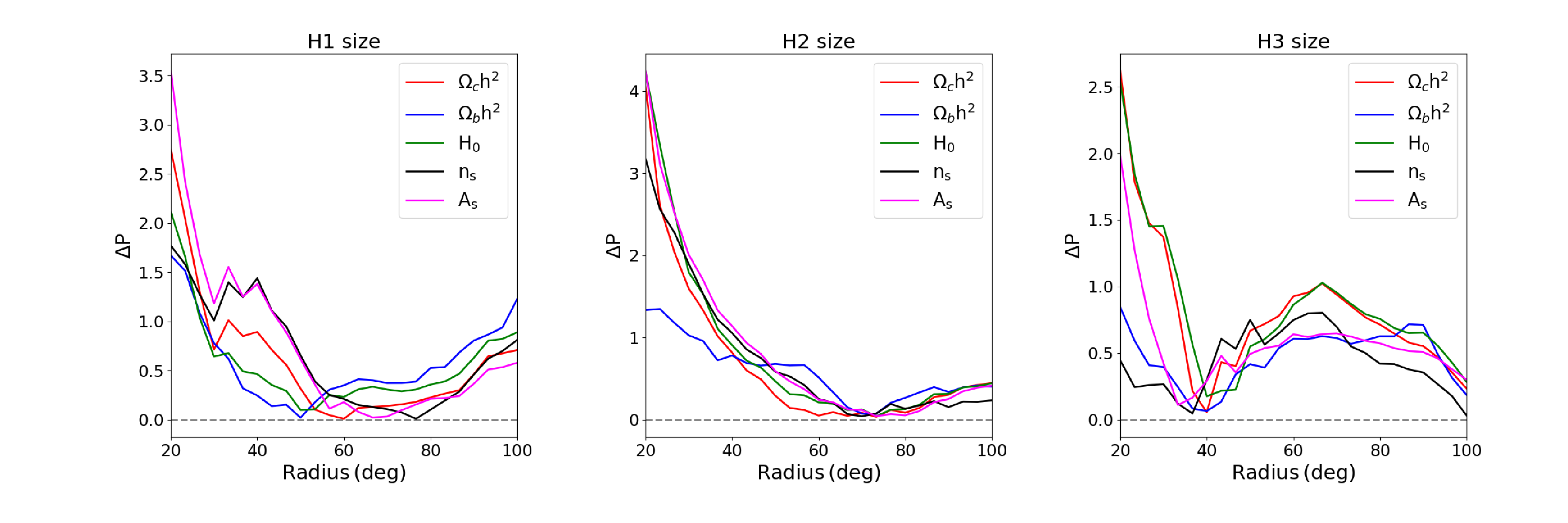}
 \caption{Same as Figure \ref{fig:hsize20} but when using discs of $60$ degrees in diameter.}
 \label{fig:hsize30}
\end{figure*}

\begin{figure*}
 \includegraphics[width=\textwidth]{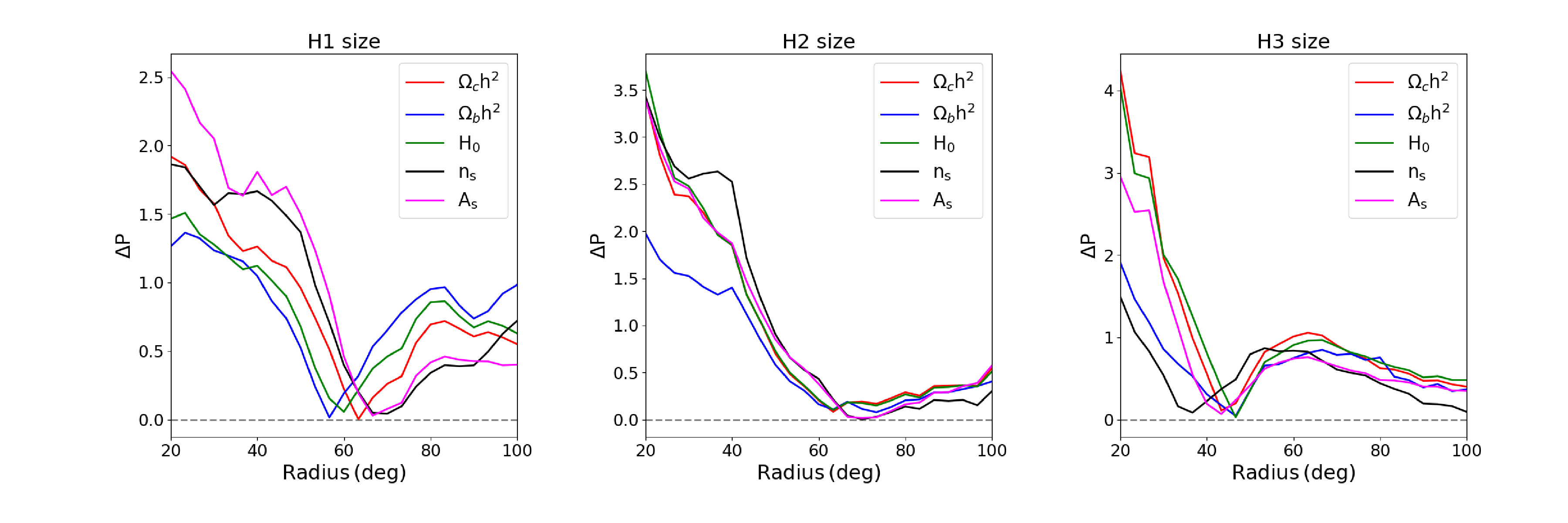}
 \caption{Same as Figure \ref{fig:hsize20} but when using discs of $90$ degrees in dimaeter.}
 \label{fig:hsize45}
\end{figure*}

\begin{figure}
 \includegraphics[width=0.48\textwidth]{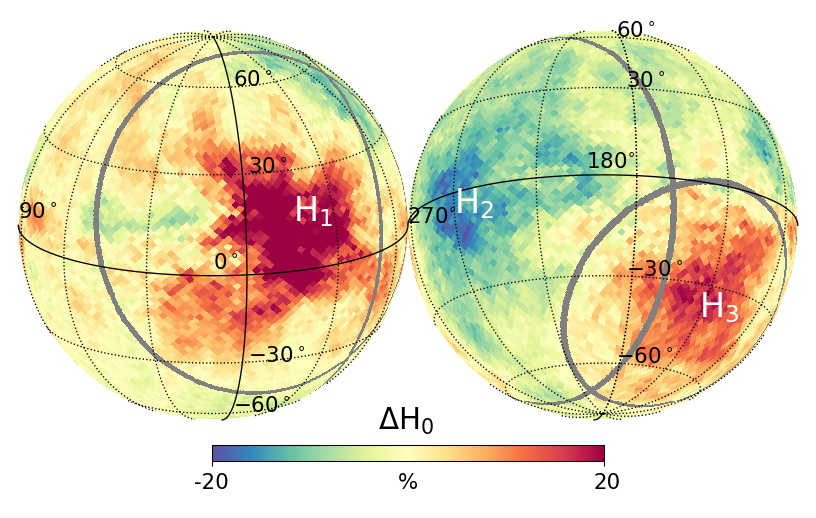}
 \caption{The three horizons (denoted by $\rm{H_1}$, $\rm{H_2}$ and $\rm{H_3}$) identified across the Hubble parameter variation map. The circular boundary of each horizon is set when the parameter variation vanishes, i.e, when it is equal to the all-sky mean (in light yellow for the color scale used). Note that the horizons partially overlap between them. Very similar horizons are obtained for the other $\Lambda$CDM parameters.}
 \label{fig:horizonsh}
\end{figure}

\begin{figure}
 \includegraphics[width=0.48\textwidth]{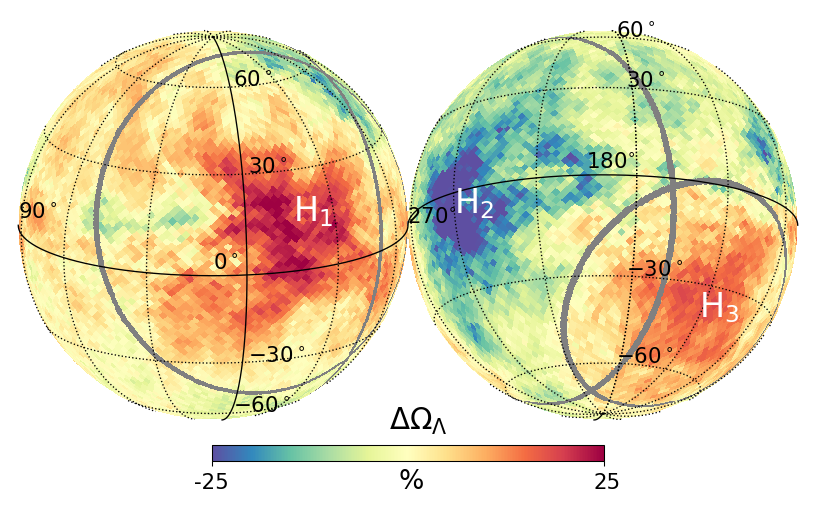}
 \caption{Same as Figure \ref{fig:horizonsh} but for the Dark-Energy density parameter.}
 \label{fig:horizonsol}
\end{figure}

We then test the robustness of the horizon size estimates across $\Lambda$CDM parameters and disc size $\cal D$ used to build the anisotropy maps. In particular, we vary the area of the discs by a factor of 5, from $40$ to $90$ degree in diameter.
Figures \ref{fig:hsize20},\ref{fig:hsize30} and \ref{fig:hsize45} show the result of using the procedure outlined above yields rather consistent results across cosmological parameters, and disc sizes. The estimated horizon size varies for the three different horizons. We estimate the sizes to be $\rm{D(H_1)  \approx 60 \pm 20}$, $\rm{D(H_2)  \approx 70 \pm 10}$, and $\rm{D(H_3)  \approx 40 \pm 10}$ degrees in radius, where the errors depict the variation across the different $\Lambda$CDM parameters (for a given disc size used) and scatter across disc sizes (for a given cosmological parameter). In Figures \ref{fig:horizonsh} and \ref{fig:horizonsol} we show the three horizons found, estimated according to Eq.(\ref{eq:horizon}), overlaid on the $\rm{H_0}$ and $\Omega_{\rm{\Lambda}}$ (derived) parameter variation maps, respectively. Similar maps are obtained for other basic $\Lambda$CDM parameters.

Our analysis shows that the main results of this paper do not depend on the particular choice of the disc  $\cal D$ size used. We have chosen $60$ degree diameter disc as the fiducial case. This is a  compromise between sample variance (which is larger for smaller discs), and over-smoothing (which is larger for bigger discs).

\subsection{Robustness to scale cuts}
\label{sec:cuts}

\begin{figure*}
 \includegraphics[width=\textwidth]{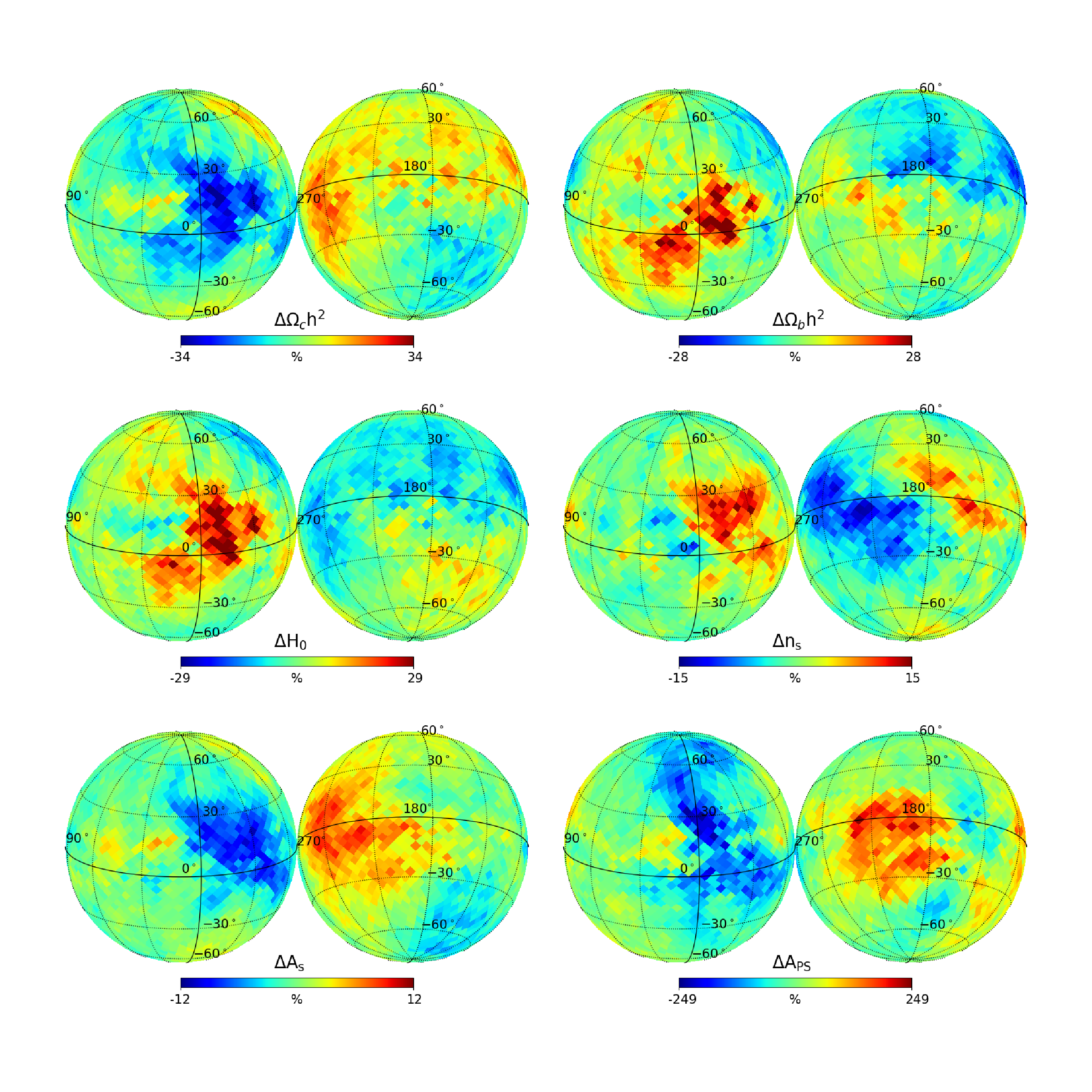}
 \caption{Same as Fig.\ref{fig:SMICAmaps}, but for the multipole range ($\ell_{min},\ell_{max}) = (450,2000)$, instead of the baseline, ($\ell_{min},\ell_{max}) = (32,2000)$. Note also that in these parameter maps, as for all the robustness tests, we use a factor of 4 coarser sampling of the sphere, i.e., $3072$ discs (instead of the $12288$ used in the baseline analysis), as this does not change our results.}
 \label{fig:horizonslmin450}
\end{figure*}

Here we discuss to what extent our results and, in particular, the main features of the cosmological parameters variation maps, are robust to the choice of scales included in the analysis. Figure \ref{fig:horizonslmin450} shows the case when we leave out the first acoustic peak from the analysis, $\ell_{min} = 450$. Even though we leave out the highest signal-to-noise contribution to the angular power spectrum, the same pattern of parameter variations identified in the fiducial case, shown in Figure \ref{fig:SMICAmaps}, remain in the maps. This points to the fact that these main features are contributed by all the multipoles of the power spectrum, and thus is not a signal coming from the largest scales alone. In turn, when we cut out signal-dominated scales the rms fluctuation of the parameter maps, $\sigma$, increases, as the relative contribution of noise is higher, and thus parameter fluctuations are artificially boosted for the same dynamic range (-4 $\sigma$, 4 $\sigma$) as can be appreciated by comparing the dynamic range of Figure \ref{fig:horizonslmin450} to the fiducal case, Figure \ref{fig:SMICAmaps}.

\begin{figure*}
 \includegraphics[width=\textwidth]{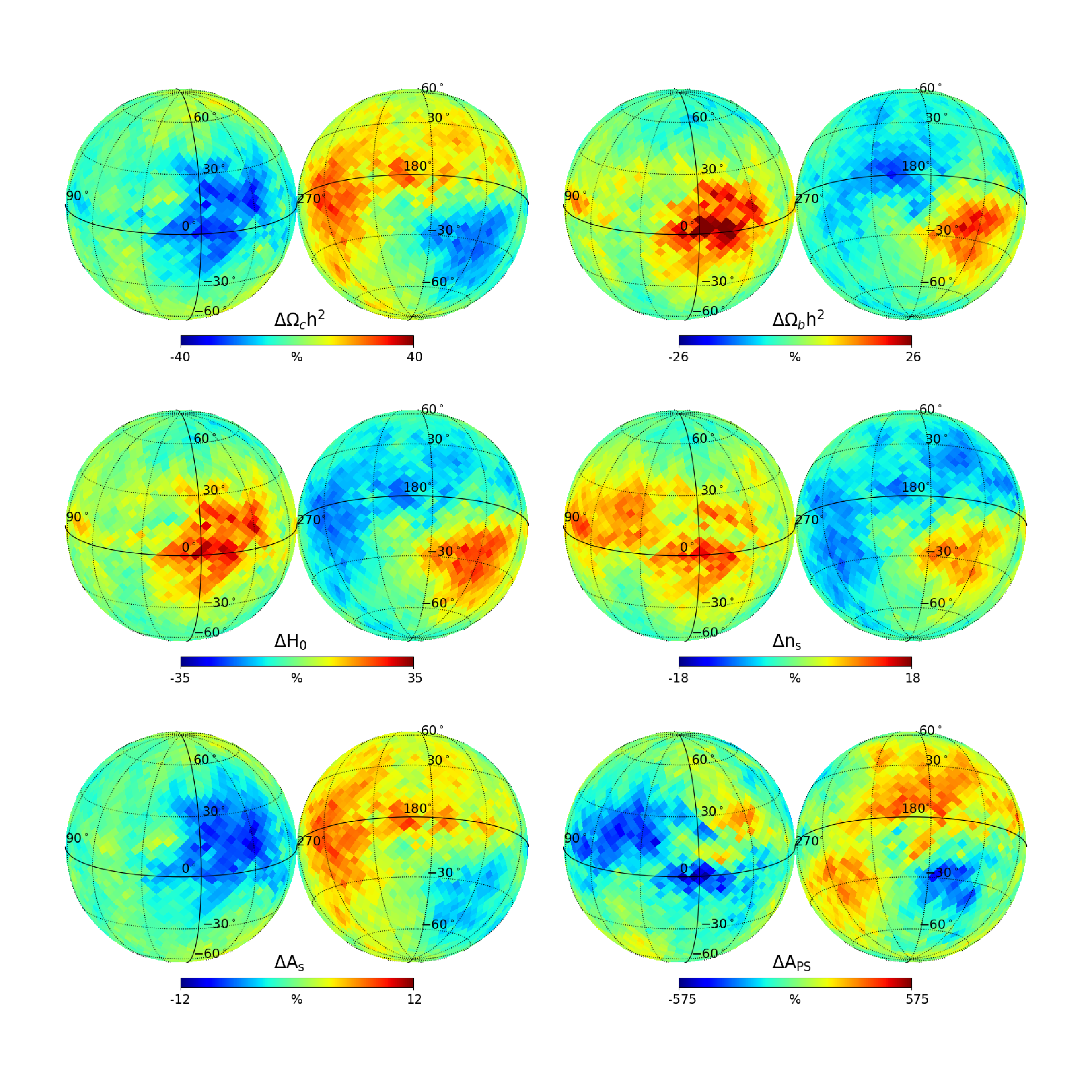}
 \caption{Same as Fig.\ref{fig:SMICAmaps}, but for the multipole range ($\ell_{min},\ell_{max}) = (32,1500)$.}
  \label{fig:horizonslmax1500}
\end{figure*}

A similar qualitative behaviour is observed for the case when we cut out multipoles beyond $\ell_{max} = 1500$ (see Figure \ref{fig:horizonslmax1500}). The slightly larger contribution from noise introduces some additional fluctuations in the parameter maps, although to a level that is significantly smaller than in the case with $\ell_{min}=450$, what does not alter the main parameter variation features across the sky. It is interesting to note that the range of scales $\ell > 1500$ in the power spectrum is the most sensitive to noise inhomogenities due to the particular scanning strategy of the Planck satellite (i.e, sweeping great circles in the sky crossing at the ecliptic poles) which, if not taken into account properly, could potentially bias our results. However, the robustness of our parameter maps to this high-$\ell$ range shows that such inhomogeneities do not contribute significantly to these maps. On the other hand, the residual foreground parameter, $\rm{A_{PS}}$, is only contributing to the power spectrum at high multipoles ($\ell > 900$, see Figure \ref{fig:clscompmap}), so the scale cut $\ell_{max} = 1500$ strongly affects this parameter map, as shown in the lower right panel of Figure \ref{fig:horizonslmax1500}.

\begin{figure*}
 \includegraphics[width=\textwidth]{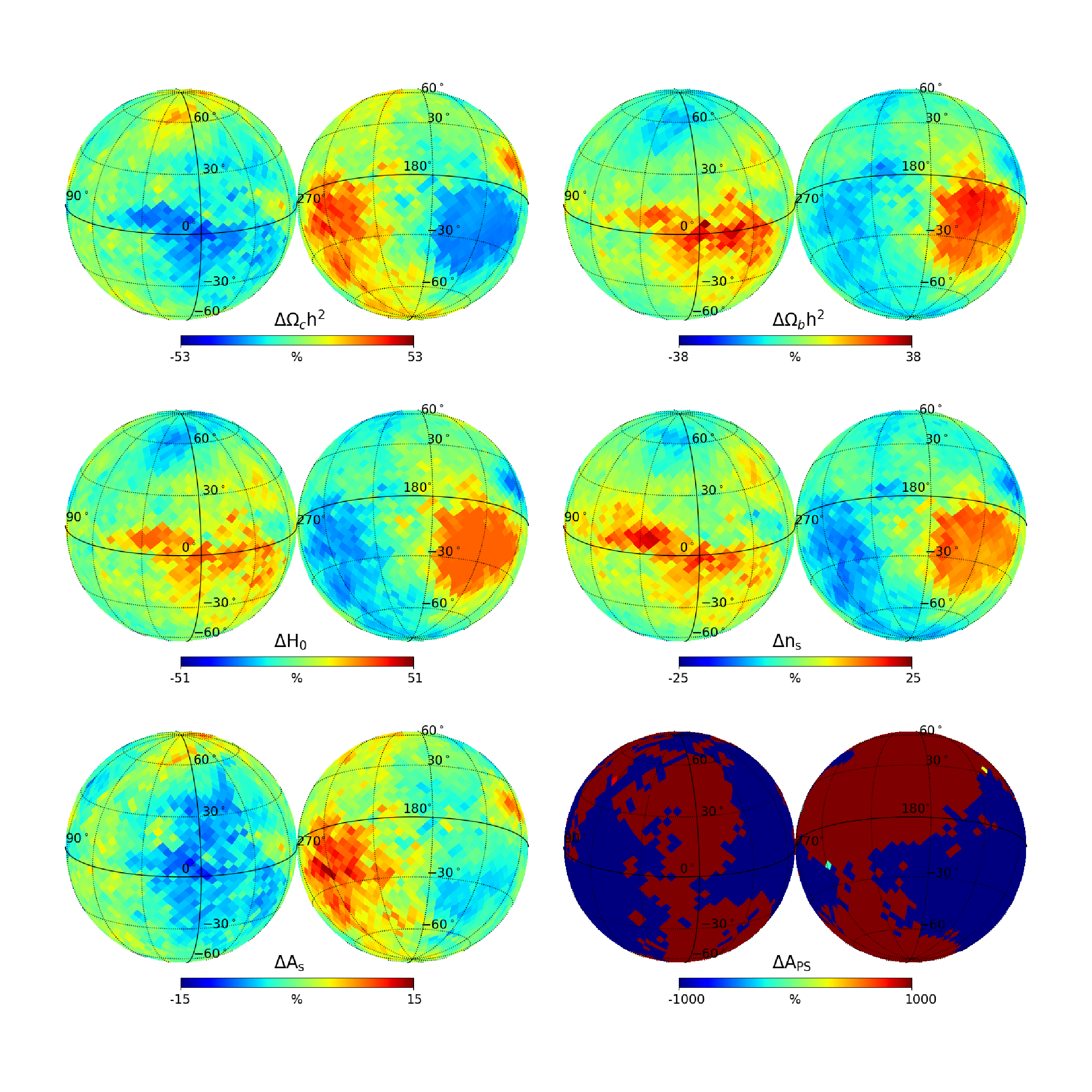}
 \caption{Same as Fig.\ref{fig:SMICAmaps}, but for the multipole range ($\ell_{min},\ell_{max}) = (32,900)$.}
  \label{fig:horizonslmax900}
\end{figure*}

Alternatively, as shown in Fig.\ref{fig:horizonslmax900}, when only the first three acoustic peaks are included in the analysis, $\ell_{max} = 900$, the main features of the parameter maps remain largely unchanged, although we start seeing some differences with respect to the full analysis (see Fig.\ref{fig:SMICAmaps}) that includes up to six peaks for $\ell_{max} = 2000$). This is to be expected, as the signal-to-noise in the angular power spectrum is well spread throughout the full dynamic range spanning from $\ell_{min} = 32$ to $\ell_{max} = 2000$, and cutting out scales beyond $\ell_{max} = 900$ leaves only about half of the total signal-to-noise available to the Planck temperature anisotropy maps and, therefore, its associated constraining power in terms of cosmological parameters is also reduced by a similar amount (see \citealt{P18cosmo}). In turn, the residual foregrounds are not well constrained in this multipole range, so the associated parameter map shows arbitrary large variations across the sky (see bottom right panel of Figure \ref{fig:horizonslmax900}).

\subsection{Robustness to foregrounds}
\label{sec:foregrounds}

\begin{figure*}
 \includegraphics[width=\textwidth]{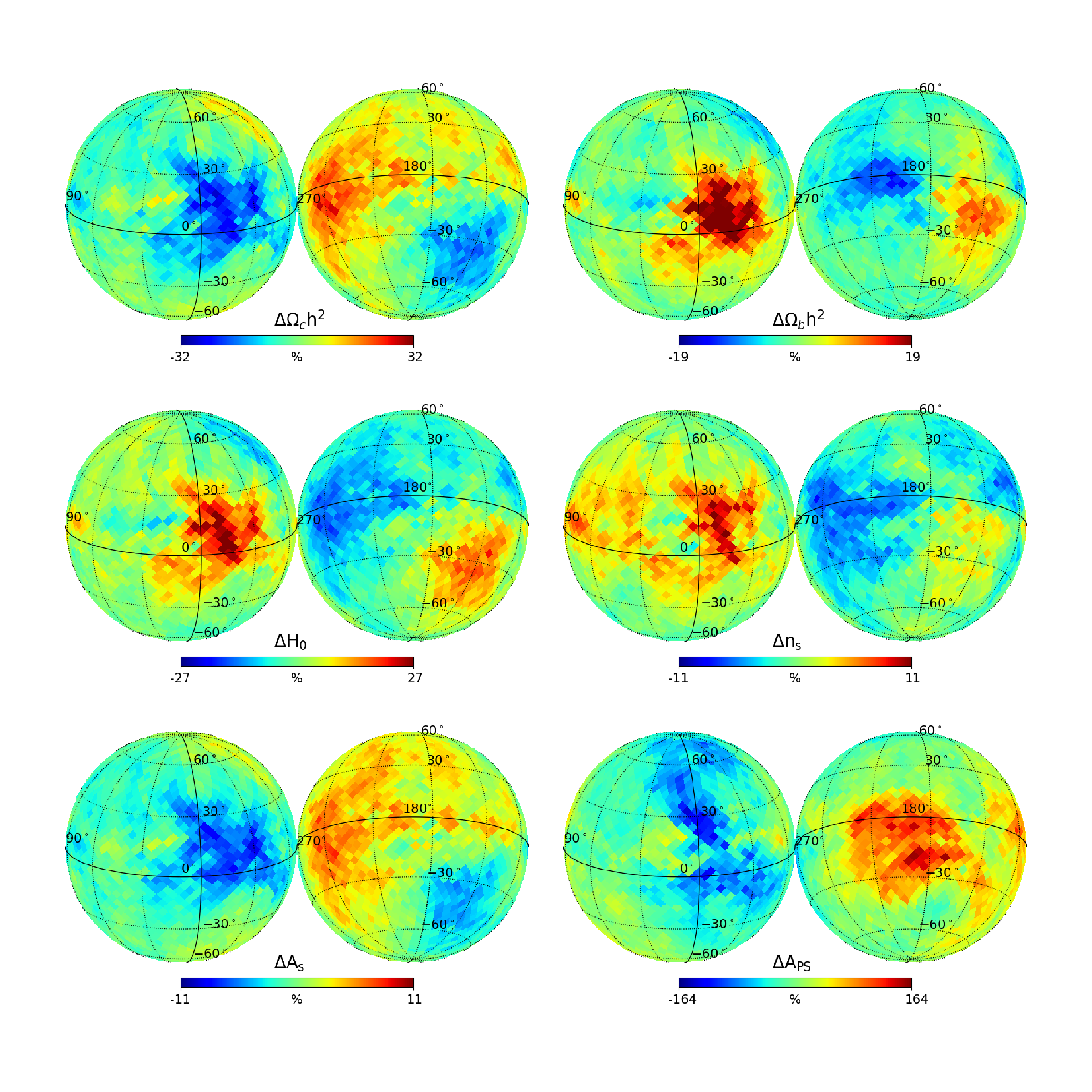}
 \caption{$\Lambda$CDM cosmological parameter variations across the Commander CMB temperature map.}
 \label{fig:CMmaps}
\end{figure*}

\begin{figure*}
 \includegraphics[width=\textwidth]{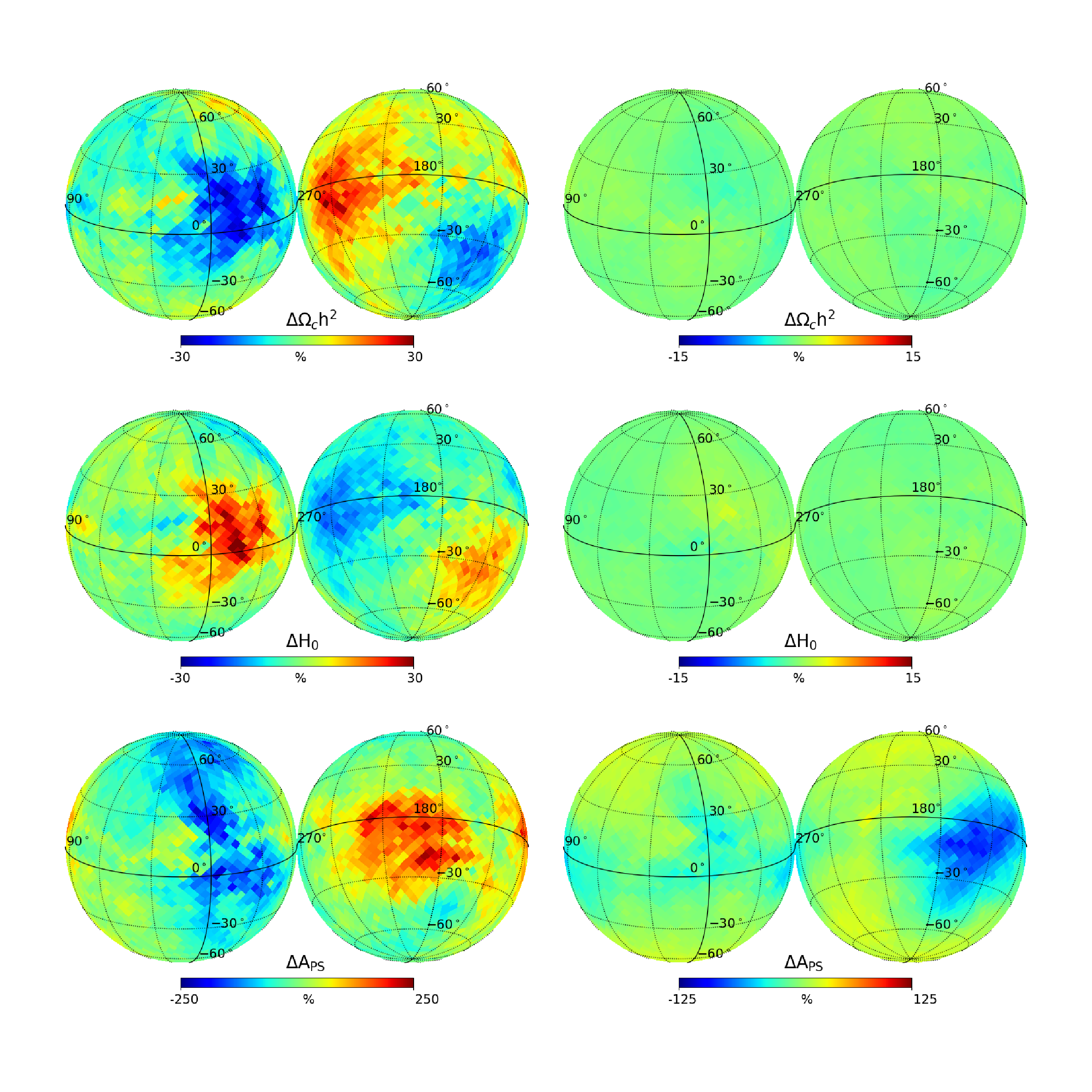}
 \caption{Half-sum (left) and half-difference (right) maps for parameter variations obtained from the SMICA vs. SEVEM  CMB maps. Top and middle panels show two examples of cosmological parameters, while the bottom panels show the case for the nuisance or residual foreground parameter.}
 \label{fig:mapcomp}
\end{figure*}

Since horizons happen to be close to the Galactic plane, it is natural to wonder whether these are affected or even produced by residual Galactic contamination. In order to address this question we have applied the same map-making pipeline for the 4 different component separation maps produced by the Planck collaboration \citep{Plegacy}: SMICA, Commander, SEVEM and NILC. These methods cover very efficient and completely independent approaches to foreground subtraction. In Figure \ref{fig:CMmaps}, we show the the case for the Commander maps, but similar maps are obtained for SEVEM and NILC. In particular, it is clear from this analysis that the cosmological parameter fluctuations across the sky are robust to component separation methods as the features in those are largely invariant across methods. On the other hand, the residual foreground component, $\rm{A_{PS}}$ exhibits large amplitude (ie., dynamic range) and spatial fluctuations across methods, suggesting that indeed the only impact of foregrounds on our analysis is through the "effective" residual point-source-like component, $\rm{A_{PS}}$. This is explicitly shown in Fig.\ref{fig:mapcomp}, where we illustrate, in the top and middle panels, how the CDM density, $\Omega_c \rm{h^2}$ and Hubble parameter, $\rm{H_0}$, display the same features in the half-sum maps, (SMICA + SEVEM/2), whereas the half-difference, (SMICA-SEVEM)/2, is consistent with no signal (note the dynamic range of color scale for the latter is half of that used in the half-sum, to emphasize residuals). This is in contrast with the case for the residual foreground parameter, $\rm{A_{PS}}$, where notable differences are seen from both component separation methods (bottom panel).

\begin{figure}
 \includegraphics[width=0.48\textwidth]{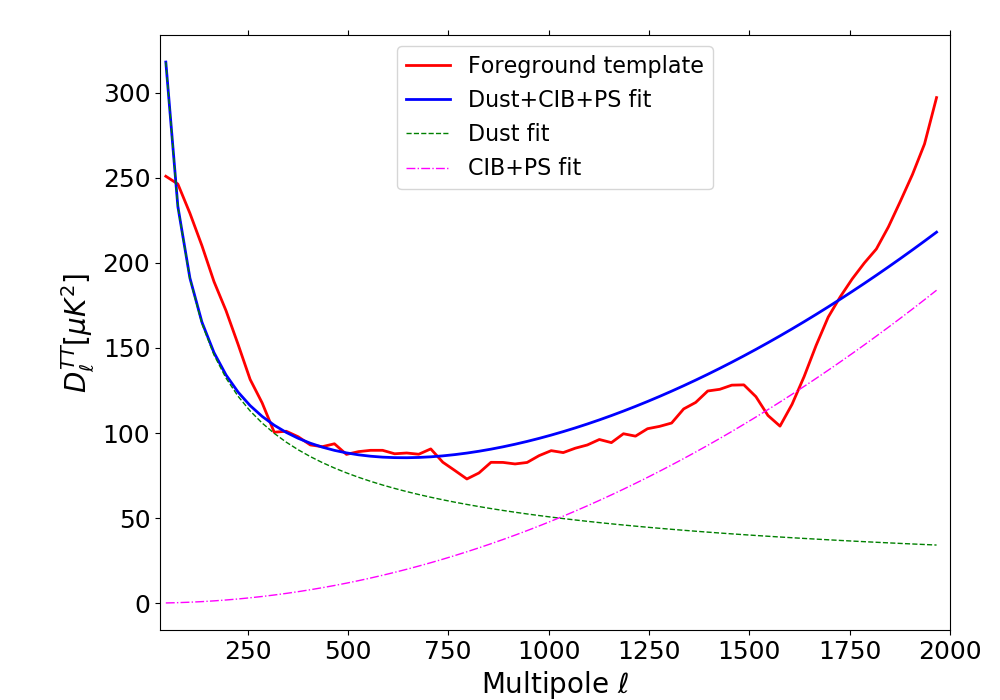}
 \caption{SMICA "CMB-subtracted" foreground template map power spectrum (red line) compared to an analytic fit (blue line) including a dominant diffuse dust emission at low multipoles (green) and an effective point-source component at high multipoles (magenta).}
 \label{fig:clscompmap}
\end{figure}

As a further test, we have investigated whether our assumptions about the foreground model template may impact our results. In particular, we have extended our single nuisance parameter in our fiducial analysis, $\rm{A_{PS}}$, that is only important at large multipoles of the CMB temperature power spectrum, to include an additional component, related to dust, at low multipoles. In order to implement this, we have derived the power spectrum of the "CMB-subtracted" SMICA foreground component separation map for the $143$ GHz channel \footnote{Maps are available at the Planck Legacy Archive \texttt{http://pla.esac.esa.int}}, since this represents a good balance between the different HFI frequency channels contributing to the CMB maps. Figure \ref{fig:clscompmap} shows the power spectrum measured from the template compared to an analytic fit that accurately reproduces the qualitative behavior of the template at low and high multipoles \citep{2016A&A...594A..11P}):
\begin{equation}
    C_{\ell} = A_{Dust} (\ell/\ell_{p})^{\alpha_{Dust}} + A_{PS}
\end{equation}
where ${\ell_p}=3000$ is a pivot or reference multipole, $\rm{A_{Dust}}$ and $\rm{\alpha_{Dust}}$ describe the Galactic dust emission amplitude and spectral index, respectively, and the shot-noise-like amplitude $\rm{A_{PS}}$ effectively encodes the CIB and extragalactic point-source contamination. We note that, in this extended foreground parametrization, we do not expect $\rm{A{PS}}$ to have the same value than in the simpler parametrization (ie., when we set $\rm{A_{Dust}}=0$). 

Using this parametrization we find the following best-fit parameters over the full Planck footprint (ie., same mask than for the SMICA temperature CMB map), $\rm{A_{Dust}} = 26.5$, $\alpha_{Dust} = -2.6$ and $\rm{A_{PS}} = 430$. This is in good agreement with the template values found by the Planck collaboration \citep{2016A&A...594A..11P}.

\begin{figure*}
 \includegraphics[width=\textwidth]{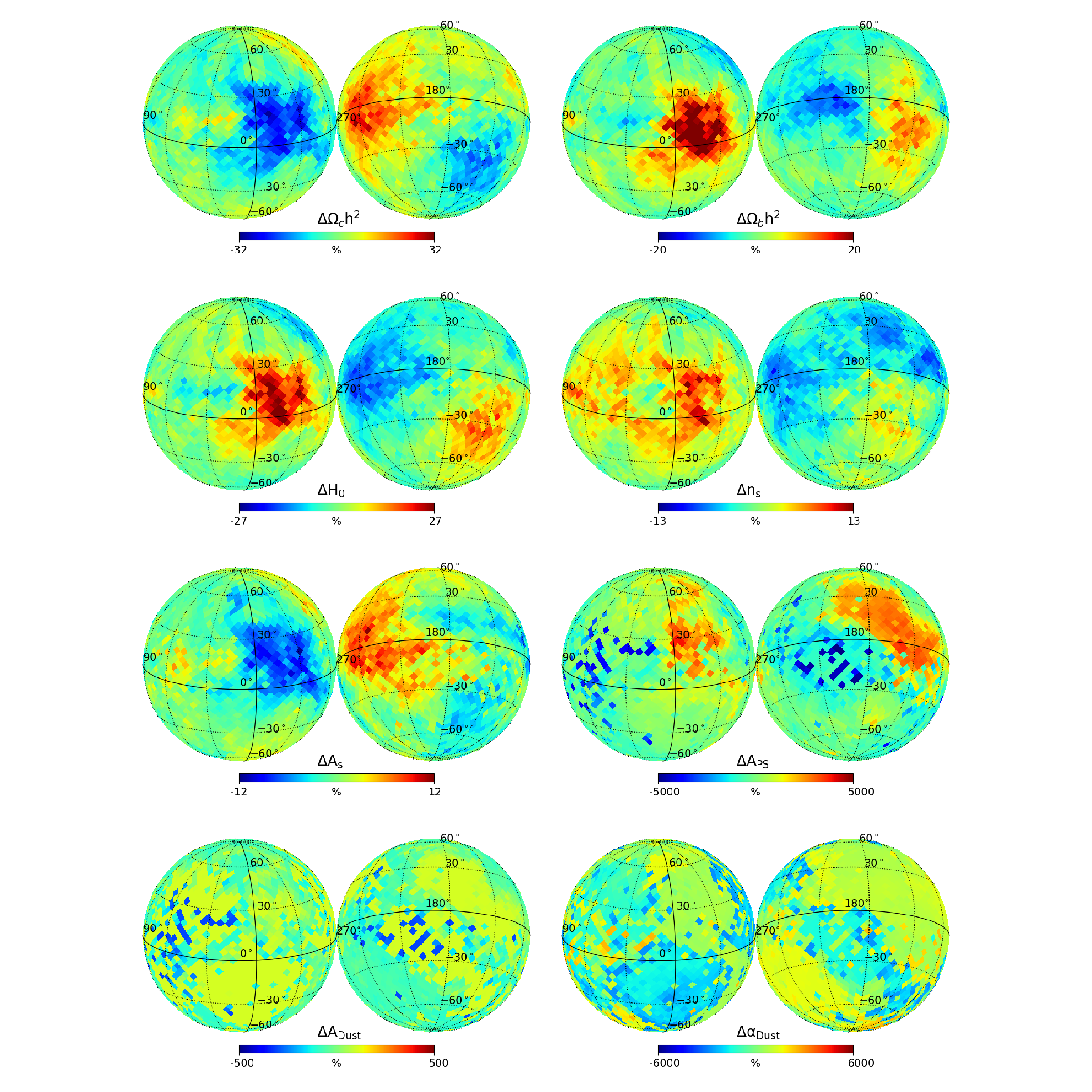}
 \caption{Parameter variation maps for the extended foreground model. Note that the maps for the $\Lambda$CDM parameters do not significantly change with respect to the single nuisance parameter case used in the fiducial analysis. The mean amplitude of the dust emission foreground parameters ($\rm{A_{dust}}$) across the sky is consistent with zero, what induces large angular fluctuations of the extended foreground parameters (see text for details).}
 \label{fig:parammaps8d}
\end{figure*}

We have used this extended nuisance or foreground parametrization in our CMB map best-fits to see if that makes any difference in our main results. In particular, we have first performance a likelihood estimation on an 8-dimensional parameter space, including the same 5 $\Lambda$CDM parameters, {$\Omega_c\rm{h^2}, \Omega_b\rm{h^2}, \rm{H_0}, \rm{n_s}, \rm{A_s}$} and the 3 nuisance parameters, $\rm{A_{Dust}}$, $\rm{\alpha_{Dust}}$, and $\rm{A_{PS}}$. When fitting this model to the Planck CMB temperature map over the full footprint, we find a vanishing dust amplitude, $\rm{A_{Dust}}=0$, what shows that our single nuisance quantity is enough to capture foreground contamination on the scale of the full footprint. Next we have performed the same test but over the average power spectra for disc-shaped patches of 60 degrees in diameter, ${\cal D}_{60}$, we used to estimate parameter variations across the sky (see
\S\ref{sec:results}), finding again no evidence for dust in the parameter fits either. As a final test, we studied whether this parametrization changes the evidence for horizons in the parameter variation maps. Figure \ref{fig:parammaps8d} shows that the features of the cosmological parameter maps, as well as the corresponding horizon locations and amplitudes are largely unaffected by this potential dust component, the only change being the estimated spatial variations of the nuisance parameter $\rm_{A_{PS}}$. The rather large and very noisy variations observed for the additional foreground-related parameters ($\rm{A_{Dust}}$ and $\rm{\alpha_{Dust}}$) suggest these are purely Gaussian sample variance fluctuations of otherwise vanishing mean quantities over the sky. Therefore this analysis suggests that dust emission does not seem to have any significant effect on the main anisotropies of the cosmological parameter maps.

\begin{figure*}
 \includegraphics[width=\textwidth]{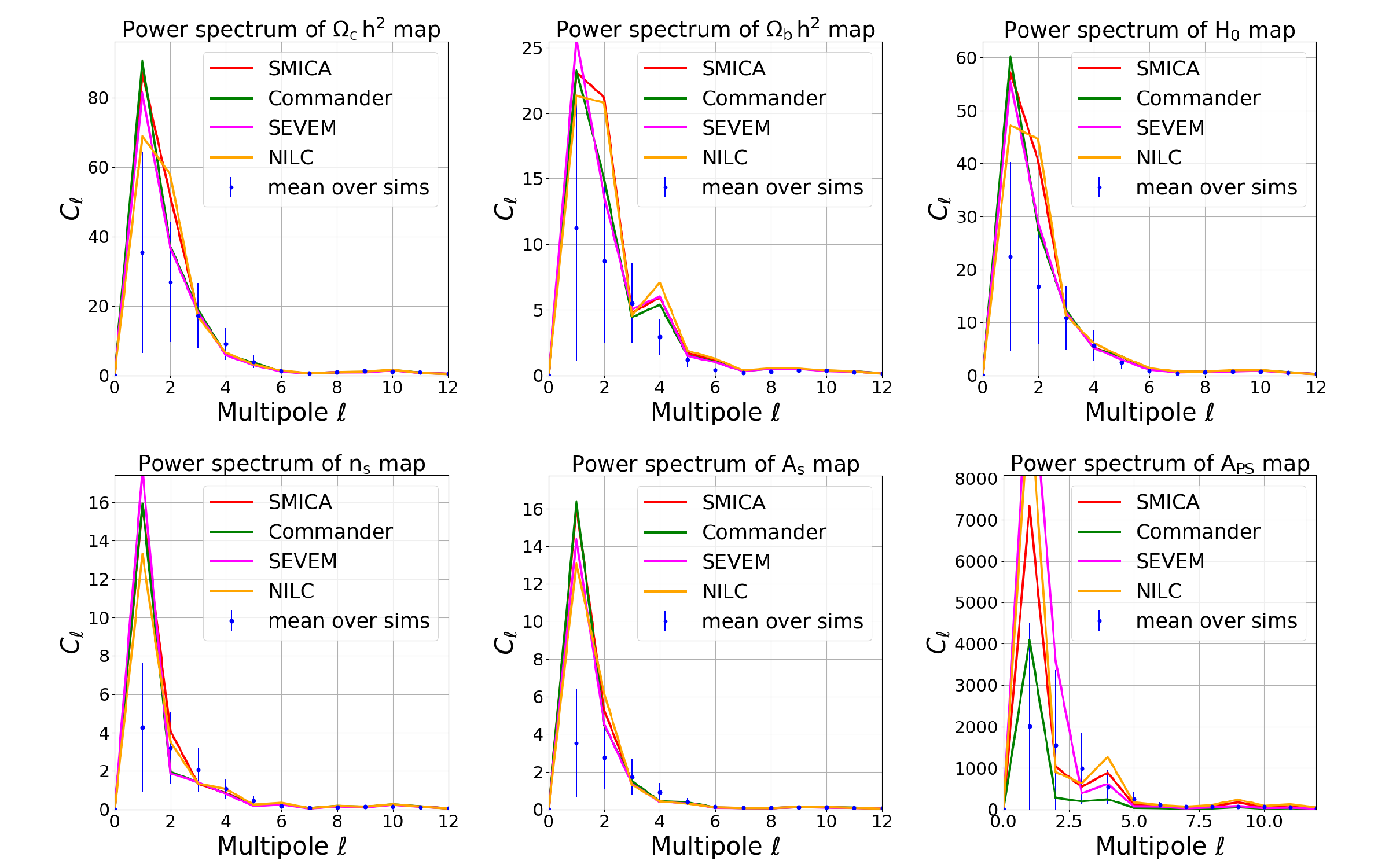}
 \caption{Power spectra of the parameter variation maps: anisotropy features in the map are encoded in lowest multipoles ($\ell < 10)$. Colors show different choices for the foreground component separation algorithms used to obtain the CMB temperature map, whereas the symbols with errorbars display the corresponding measurements from Gaussian isotropic simulations (see section \S\ref{sec:hsignificance} for the related discussion of the significance of the power spectrum monopole and dipole of the parameter maps that use as input the fiducial (SMICA) foreground cleaned map.}
 \label{fig:clsparams}
\end{figure*}

Moreover, we can statistically quantify the possible impact of foregrounds (mainly dust, CIB, and extragalactic point sources) on the features that we observe in the cosmological parameter maps. For this purpose, we have studied the possible dependence of the $C_{\ell}$'s of the parameter maps on the different available versions of the foreground cleaned temperature anisotropy maps. Accordingly, we have replaced the SMICA map with the other component separation maps provided by the Planck collaboration, ie. the Commander, SEVEM and NILC  maps.  Figure \ref{fig:clsparams} shows that our results are robust to different foreground cleaning methods. In particular, only few $\%$ variations in the dipole and quadrupole amplitudes are observed, but largely keeping the sum of them unchanged. Thus we conclude that the "horizon"-like features of the parameter maps, which are encoded in such dipole and quadrupole moments, are not significantly affected by foreground emission. 

\subsection{Comparison to WMAP}
\label{sec:wmap}

If the $\Lambda$CDM best-fit parameter variations across the CMB sky are truely cosmological in origin, one should we able to detect them using any CMB dataset, provided it has enough signal to noise.
In turn this provides an important test for survey specific systematics as a possible source for the detected anisotropic signal. As a cross-check of our main results, based on the Planck temperature data maps, we shall perform an equivalent analysis on the WMAP 9-year data \citep{WMAP9}, which has a lower angular resolution and higher noise per power spectrum multipole beyond the first acoustic peak with respect to Planck. However, we expect to find evidence for consistent patterns of angular variations of the best-fit cosmological parameters across the sky for WMAP, although at lower statistical significance. 

\begin{figure}
 \includegraphics[width=0.48\textwidth]{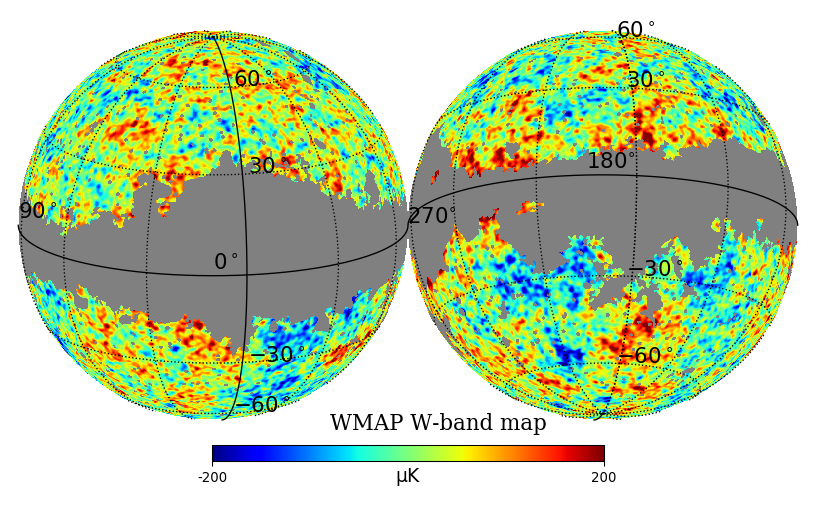}
 \caption{WMAP temperature anisotropy map for the W frequency band, smoothed with 1 degree FWHM Gaussian beam. The Galactic mask used is overlaid.}
  \label{fig:wmap_sm1deg}
\end{figure}

Our WMAP data analysis pipeline is described below:
\begin{itemize}

\item {\it Step 1: Data}. Following \cite{WMAP9}, we use the final (9-year) temperature data maps from the LAMBDA website\footnote{\tt{https://lambda.gsfc.nasa.gov}} for the 6 channels within the V and W frequency bands (V1,V2,W1,W2,W3,W4). We note that we use the maps with the asymmetric component of the beam "deconvolved", so as to avoid introducing spurious anisotropic signals from CMB maps in our results (see \citealt{2010PhRvD..81j3003H}). As for the Galactic mask, we use the "Extended Temperature Analysis Mask" (KQ75y9) that leaves about $69\%$ of the sky available for the analysis. Fig.\ref{fig:wmap_sm1deg} shows an example map (W-band frequency channel) smoothed with a 1 degree Gaussian beam (for better visualization) along with the mask used
\item {\it Step 2: Power spectrum estimation per channel}. We compute angular power spectra for each pair, with an inverse variance (inhomogeneous) noise weighting given by $w = \sqrt{N_{obs}}/\sigma_0$, where $N_{obs}$ is the number of observations per pixel and $\sigma_0$ is the noise per observation for a given channel (as given in Table 5 of \citealt{WMAP9}), and deconvolve the resulting $C_{\ell}$'s with the (symmetrized) beam transfer functions as provided in the LAMBDA website. We then subtract an estimate of the frequency-dependent bias due to unresolved radio point sources (see Section \S 6.2 in \citealt{WMAP9})
\item {\it Step 3: Final Power Spectrum}. The final power spectrum is obtained by combining the cross-spectra from the 15 channel pairs using an inverse noise weighting given by $w_{i,j} = 1/\sqrt{\sigma^i_0\sigma^j_0}$, where the indexes $i,j$ run over all the channel pairs and $\sigma^i_0$ is the noise per observation for the $i$ channel
\item {\it Step 4: Multipole range for likelihood estimation}. In order to avoid potential foreground contamination, we set a lower multipole $\ell_{min} = 32$ as we did for Planck, and a maximum multipole $\ell_{max} = 900$ to avoid beam suppresion effects and residual high multipole foregrounds in the analysis. This in turn means that we only need to fit the 5 $\Lambda$ CDM cosmological parameters to the final power spectra as the residual foreground amplitude $\rm{A_{PS}}$ is negligible on these scales (unlike the reference analysis we presented above for Planck)

\end{itemize}

\begin{figure}
 \includegraphics[width=0.48\textwidth]{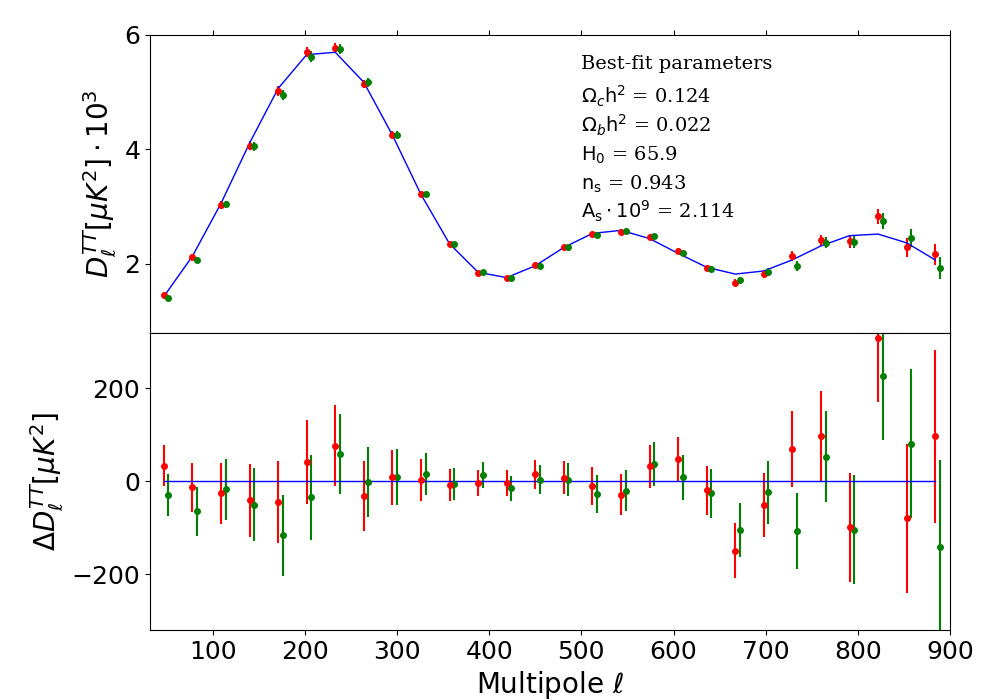}
 \caption{Estimate of the angular power spectrum of the WMAP temperature map (red symbols), compared to the published results from the WMAP team (green symbols, slightly shifted to the right for clarity). The spectrum for the best-fit parameters is shown by the solid blue line. Lower panel shows residuals of the estimated spectrum with respect to the best-fit.}
  \label{fig:wmapcls}
\end{figure}

As a validation of the above analysis pipeline, Figure \ref{fig:wmapcls} shows the resulting angular power spectrum (and derived best-fit $\Lambda$CDM parameters), when we use the full available area from the WMAP mask, compared to the WMAP team results \citep{WMAP9}, showing very good agreement, given the errors.

\begin{figure*}
 \includegraphics[width=\textwidth]{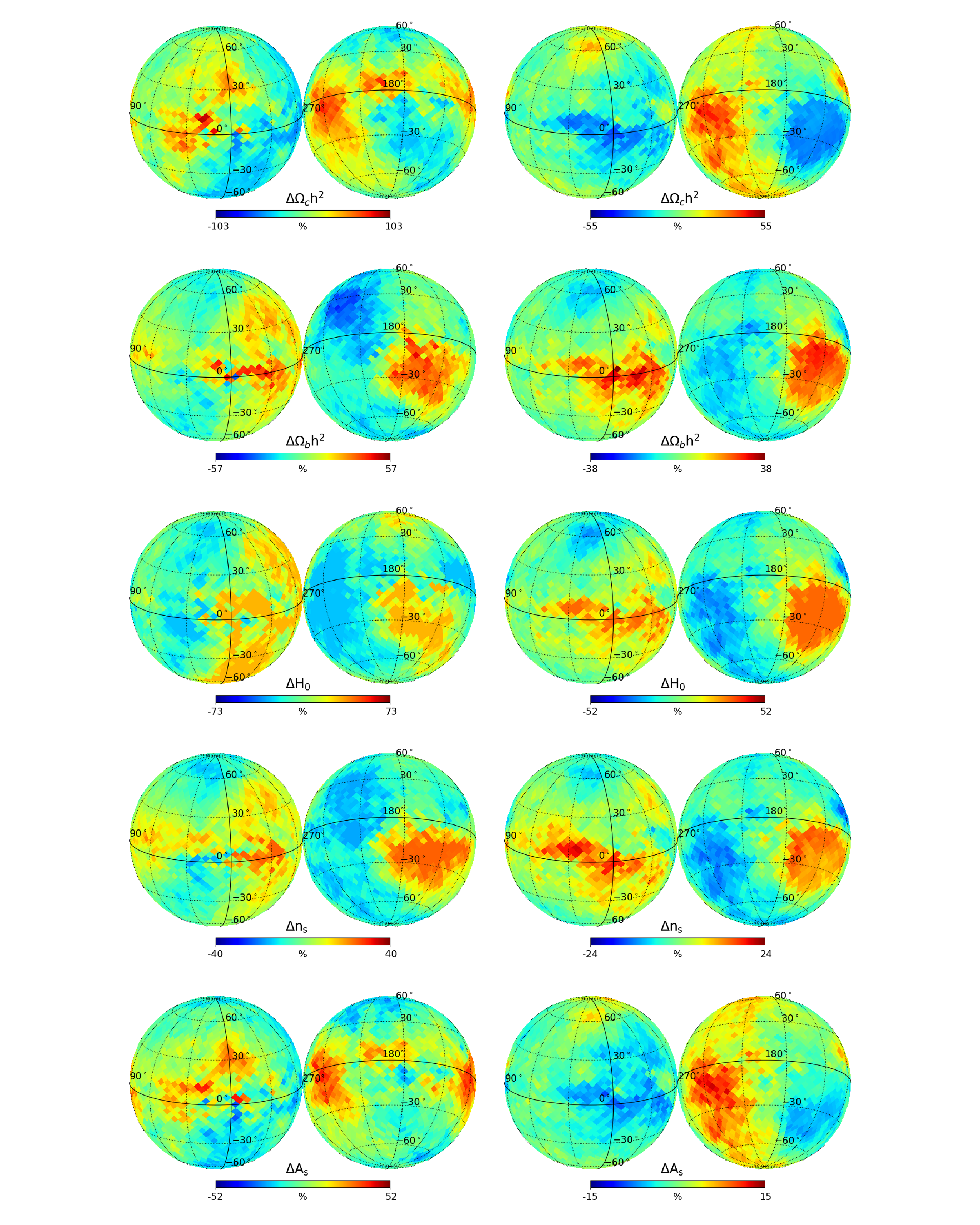}
 \caption{$\Lambda$CDM parameter maps for WMAP (left) and Planck (right). Both datasets use the same scale cuts, as detailed in the text.}
  \label{fig:wmapVSplanck}
\end{figure*}

We then repeat this analysis but for each of the $3072$ discs of $60$ degree diameter distributed across the sky. The resulting best-fit parameters for each disc are shown in the left panels of Figure \ref{fig:wmapVSplanck}, whereas the corresponding results for Planck, using the same scale cuts (i.e, $\ell_{min} = 32$ and $\ell_{max} = 900$) are displayed in the right panels. Although the WMAP signal-to-noise with the used scale cuts is $~3$ times lower than the full Planck analysis (i.e, with $\ell_{max} = 2000$) we still expect to see some hints of the same anisotropic signals across WMAP maps. Remarkably, for the parameters $\Omega_b \rm{h^2}$, $\rm{H_0}$ and $\rm{n_s}$, we do observe the same large scale anisotropic patterns (i.e, horizons) than in Planck, whereas for $\Omega_c \rm{h^2}$ and $\rm{A_S}$, we find some differences, specially in the left hemisphere (i.e, the one around the galactic center). in fact, some differences are to be expected since, as we mentioned, WMAP data has larger noise  (and lower resolution) than Planck.
In fact, for WMAP we expect the noise to contribute to a similar (although somewhat lower) level than the signal to the parameter maps, what seems to be in agreement with the fact that larger parameter variations are observed in the WMAP maps, as compared to Planck.

\subsection{Horizons as the source of power asymmetry}
\label{sec:powerasym}

The basic features of acoustic peaks in the CMB power spectra can be entirely determined from three fundamental scales or multipoles that depend on the cosmological parameters \citep{Hu01}: the matter-radiation equality ($\rm{\ell_{EQ}}$), the acoustic ($\rm{\ell_A}$), and the damping scale ($\rm{\ell_D}$). From the set of $3072$ discs of 60 degree in diameter, $\rm{{\cal D}_{60}}$, we estimate $\rm{\ell_{EQ}}=148 \pm 7$, $\rm{\ell_{A}}=304.0 \pm 0.6$, and $\rm{\ell_{D}}=2018 \pm 38$. Furthermore we can thus use the cosmological parameter variation maps measured in section \S\ref{sec:cosmoparams} to derive the corresponding variations of these scales across the sky. In particular, we use the following expressions for a flat $\Lambda$CDM model (see Eqs.(A15)-(A17) in \cite{Hu01}):
\begin{equation}
    \frac{\Delta \ell_{EQ}}{\ell_{EQ}} \approx -0.48 \ \frac{\Delta h}{h} + 0.07 \ \frac{\Delta \omega_b}{\omega_b} - 0.15 \ \frac{\Delta \omega_m}{\omega_m} \ ,
\end{equation}
\begin{equation}
    \frac{\Delta \ell_{A}}{\ell_{A}} \approx \frac{\Delta h}{h} + 0.59 \, \frac{\Delta \omega_m}{\omega_m} \ ,
\end{equation}
and,
\begin{equation}
    \frac{\Delta \ell_{D}}{\ell_{D}} \approx 0.42 \ \frac{\Delta h}{h} + 0.20 \ \frac{\Delta \omega_b}{\omega_b} - 0.12\ \frac{\Delta \omega_m}{\omega_m} \ ,
\end{equation}
being $\rm{\omega_m = \omega_c + \omega_b}$, with $\rm{\omega_c = \Omega_c h^2}$, $\rm{\omega_b = \Omega_b h^2}$, and $\rm{h =H_0/100}$.

\begin{figure}
 \includegraphics[width=0.5\textwidth]{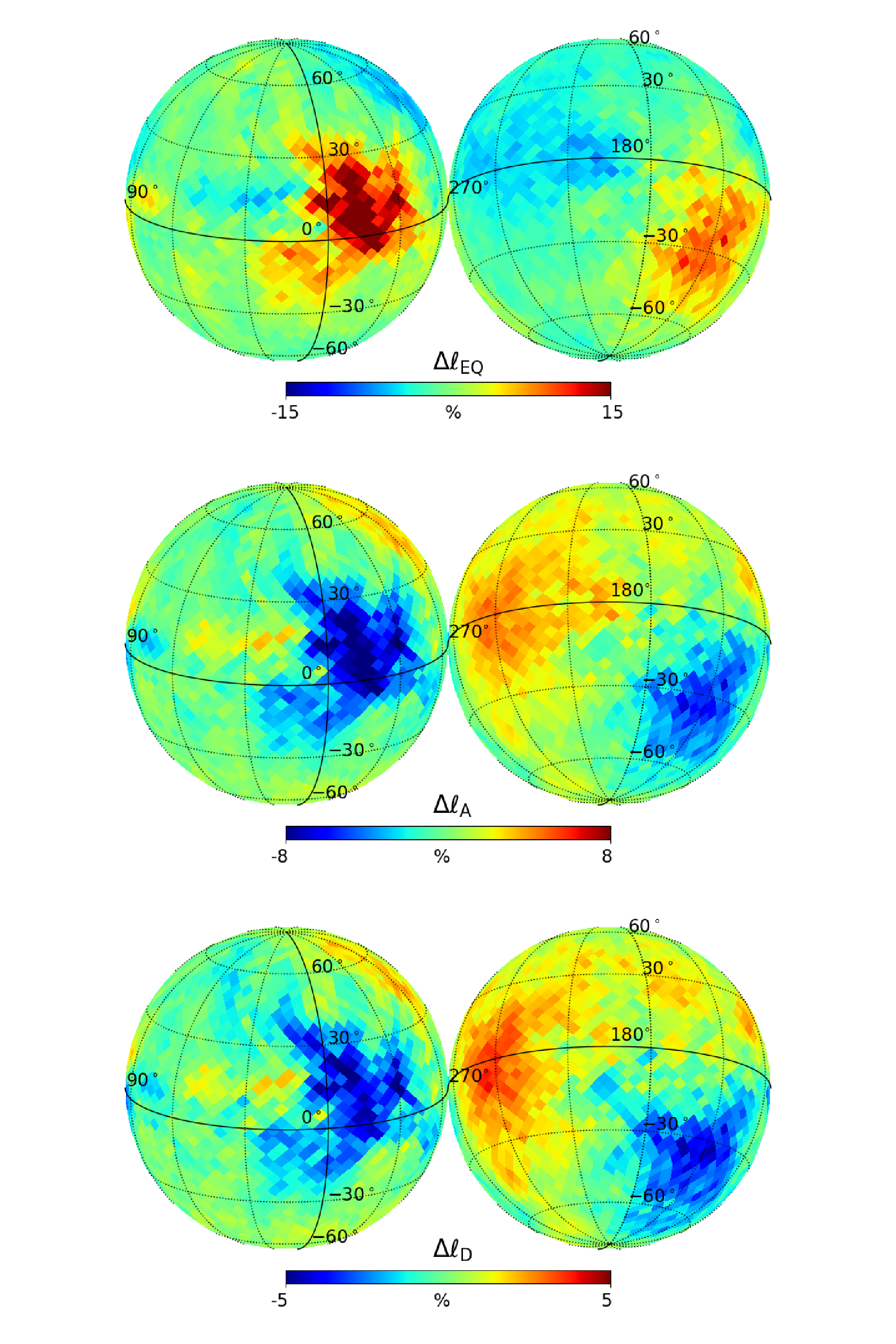}
 \caption{Maps of spatial variations of the fundamental scales of the CMB power spectrum: top panel shows changes in the equality scale, middle panel shows the corresponding map for the acoustic scale and the bottom panel the damping scale (see text for details). These maps show the same features than in the cosmological parameter maps (see e.g. Figure \ref{fig:SMICAmaps}).}
 \label{fig:fundscales}
\end{figure}

Figure \ref{fig:fundscales} shows that the relative spatial variations of these fundamental scales exhibit the same features observed for the basic $\Lambda$CDM parameter maps, displaying in particular the three coherent "horizon" shapes described above. This in turn emphasizes the universal nature of the features attributed to the cosmological horizons.

On the other hand, a long-standing and puzzling anomaly in CMB analyses is the significant hemispherical power asymmetry first found in WMAP data analyses (\citealt{H04,E07,2008JCAP...09..023L,H09,2009ApJ...699..985H,2010MNRAS.407..399P,2013ApJ...773L...3A,2013PhRvD..87l3005D}) and more recently confirmed using Planck data (\citealt{P13isotropy,Akrami14,2015PhRvD..92f3008A,P15isotropy,2016JCAP...06..042M,P18isotropy}). In particular, \cite{Akrami14} claimed a detection at the $3.3 \sigma$ level, arguing that their measurement is robust to systematic effects and foregrounds, what points to a (yet unknown) physical origin for the source of this anisotropy.

We shall provide below a simple explanation for the origin of this power asymmetry in terms of the cosmological horizons, i.e, physically disconnected regions of our universe. In particular, if these horizons do exist, as our analysis suggest (see Section \S\ref{sec:hsignificance} below for their significance), the features they imprint on the maps of $\Lambda$CDM parameters or fundamental scales could be responsible for the observed CMB power asymmetry. In previous analyses, \cite{H09} found maximal asymmetry in WMAP data for a reference direction (center of one of the hemispheres) pointing to $(l,b) = (226^{\circ},-17^{\circ})$, which is consistent with more recent estimates by \cite{Akrami14}, who estimate a preferred direction towards $(l,b) =  (212^{\circ},-13^{\circ})$ using Planck data. These directions are broadly consistent (within their estimated errors, see Fig.3 in \citealt{Akrami14}) with the center of the $\rm{H_2}$ horizon, located at $(l,b) = (240^{\circ},-5^{\circ})$, (see Table \ref{tab:hlocation}).

\begin{figure}
 \includegraphics[width=0.48\textwidth]{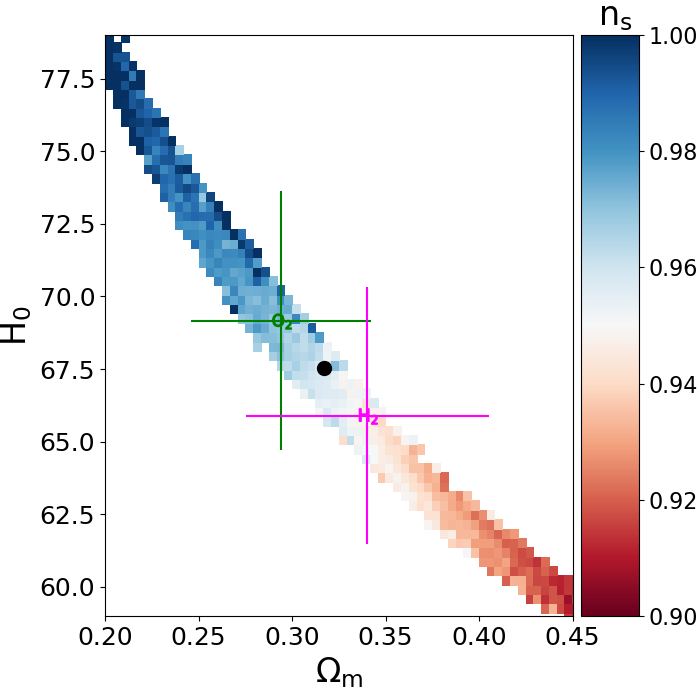}
 \caption{Power asymmetry from cosmological horizons: differences in $\rm{\Omega_m}$, $\rm{H_0}$ and $\rm{n_s}$ for the hemispheres pointing towards the center of the $\rm{H_2}$ horizon (purple symbol with errors) and its opposite direction on the sky $\rm{O_2}$ (green). The mean value across the sky is shown for reference (black dot).}
 \label{fig:powerasym}
\end{figure}

To illustrate how this particular test of cosmological anisotropy also seems to originate from the existence of such horizons, in Figure \ref{fig:powerasym} we show the (3D) distribution of values for the matter density $\rm{\Omega_m}$ and Hubble parameter $\rm{H_0}$ as a function of the spectral index $\rm{n_s}$ in discs distributed across the sky. In particular, non-negligible shifts in estimated best-fit parameters are inferred from the hemisphere centered at the $\rm{H_2}$ horizon with respect to the opposite direction (denoted by $\rm{O_2}$ in Figure \ref{fig:powerasym}).
In particular, inside the hemisphere pointing towards $\rm{H_2}$, we find a $15 \%$ higher matter density and $5 \%$ lower expansion rate, that is correlated with a $2 \%$ smaller spectral index, as compared to the values found in the opposite ($\rm{O_2}$) hemisphere. Similar albeit somewhat smaller parameter differences are found when defining hemispheres pointing to the $\rm{H_1}$ and $\rm{H_3}$ horizons. However we must emphasize that the cosmological parameter hemispherical asymmetries hinted above are not found to be statistically significant given the estimated errors (i.e, scatter across discs measurements). 

\subsection{Simulations}
\label{sec:sims}

In order to quantify the significance of the measured angular features in the cosmological parameter maps we have analyzed two sets of simulations. First, we have used a set of 300 "Full Focal Plane" (FFP10) lensed CMB maps\footnote{Planck simulations are available from the Planck Legacy Archive webpage, \texttt{http://pla.esac.esa.int} and they are described in the Planck Legacy Archive wiki,\texttt{http://wiki.cosmos.esa.int/planck-legacy-archive/index.php/Simulation\_data}}, which are convolved with the FEBeCoP effective beams. We note that we limit our analysis to 300 CMB realizations per frequency channel (out of the 1000 available in the Planck Legacy Archive) so as to match the number of corresponding FFP10 noise realizations available in the Planck Legacy Archive. CMB maps are provided for the full-mission only and, in order to follow the Planck likelihood estimation (see \cite{2020A&A...641A...5P}), we need to combine angular power spectra from the $100, 143$ and $217$ GHz frequency channels. This amounts to a total of 900 CMB simulated maps. 

On the other hand, as mentioned above, for each frequency channel, we have also used the 300 available realizations of the (anisotropic) noise and instrumental effect residual maps. We select the version corresponding to the "Odd" and "Even" (OE) stable pointing periods to model the corresponding OE data cuts used in our analysis. Again, we need to combine noise maps from the three relevant HFI frequency channels in order to properly model the Planck likelihood estimation, what amounts to a total of 900 noise realizations ($\times$ 2 data cuts). Finally we can combine each CMB and noise realization (for each frequency channel and OE data cut) to produce a set of "realistic" simulations of the Planck temperature maps.

In order to validate the CMB plus noise realizations, we make use of our validated cosmological parameter estimation pipeline described in Section \S\ref{sec:pipevalid}, involving the computation of cross power-spectra $C_{\ell}$'s using the same Galactic mask that is applied to the analysis of the real data (see Section \S\ref{sec:method}) . This involves computing 1500 cross power-spectra involving OE1$\times$OE2 rings maps for the 5 frequency pair combinations ($100\times100, 100\times143, 143\times217, 217\times143$ and $217\times217$ GHz) for each of the 300 realizations per frequency channel. 

As a first validation, we estimated the cosmological best-fit parameters from the average of the 300 simulated map power spectra using the full Planck mask (instead of the set of discs) to see if we recover the input flat-space $\rm{\Lambda CDM}$ cosmological parameters given in the Planck Legacy Archive Simulations wiki ($\Omega_c\rm{h^2} = 0.120$, $\Omega_b\rm{h^2}=0.0222$, $\rm{H_0}=67$ Km/s/Mpc, $\rm{n_s}=0.964$, $\rm{A_s\cdot 10^{9}}=2.120$)\footnote{Although we have also fixed the optical depth to reionization, $\tau = 0.060$, and the sum of the neutrino masses,$\rm{\sum m_{\nu} = 0.060 \, eV}$, to the values given in the Planck wiki, we have checked that setting them to the Planck best-fit values (see Table \ref{tab:cosmo}) does not change the outcome of our validation}.

\begin{figure}
 \includegraphics[width=0.48\textwidth]{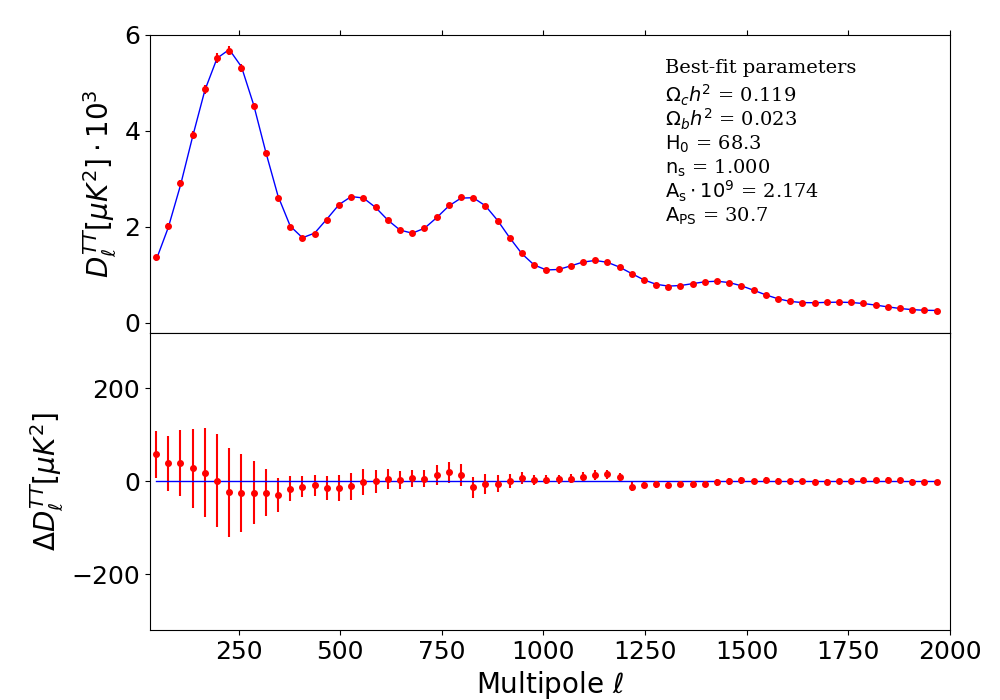}
 \caption{Mean power spectrum for a set of 300 Planck temperature FFP10 CMB+noise realizations compared to its best-fit model, i.e, basic $\Lambda$CDM cosmology plus an additional nuisance (residual foreground) parameter (see text for details). Lower panel depicts differences between simulation measurements (symbols with errorbars) and the best-fit model (solid blue line).}
 \label{fig:cmbnoisecls}
\end{figure}

\begin{figure}
 \includegraphics[width=0.48\textwidth]{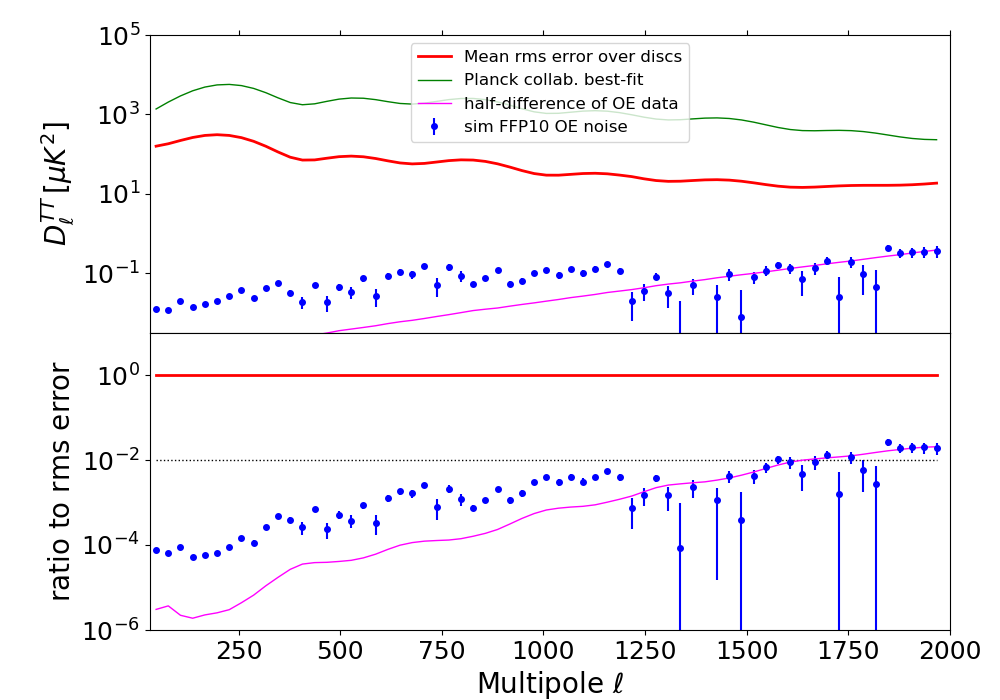}
 \caption{Impact of the anisotropic noise: the mean power spectrum for a set of 300 FFP10 noise realizations (blue symbols) compared to the mean Planck temperature power spectrum measurements (green line) and associated errors (thick red line) in discs across the sky. Lower panel shows that the average anisotropic noise amplitude (which is comparable to the simple noise estimate from the half-mission half difference maps, magenta line) is of order $1\%$ of the amplitude of the 1-$\sigma$ errors in the power spectra used to derive the cosmological parameter maps.}
 \label{fig:noisecls}
\end{figure}

Figure \ref{fig:cmbnoisecls} shows the resulting mean power spectrum estimated from the 300 FFP10 CMB plus noise realizations\footnote{We note that the small discontinuities or jumps in the band power estimates depicted at multipoles $\ell = 800, 1190$, most visible in the lower panel of Figure \ref{fig:cmbnoisecls}, reflect the slightly step-like change in the weights applied when combining the power spectra from different frequency pairs, approximately following the scheme used in the Planck likelihood estimation (see \cite{2020A&A...641A...5P})}. As shown in the figure, the best-fit cosmology is consistent with a scalar spectral index $n_s = 1.00$ (see dashed line in Figure \ref{fig:clfit} what is clearly at variance with the claimed input cosmology, with $n_s = 0.964$ , for the FFP10 CMB realizations ( see solid magenta line in Figure \ref{fig:clfit}) according to the Planck Legacy Archive Simulations wiki. We have also verified that using CMB realizations alone (i.e, without adding noise) we consistently recover the same best-fit spectral index. Moreover, the recovered value of $n_s$ is more than 6 $\sigma$ away from the Planck best-fit value (see Figure \ref{fig:clfit}, and Table \ref{tab:cosmo}), what makes the cosmology simulated rather unrealistic. Therefore it is clear that something is wrong
with this set of simulations and we have not been able to clarify exactly what. In view of this, we decided not to use the FFP10 CMB simulations for our analysis, since we could not validate them with the same pipeline we used for the data and with which we could recover the best-fit values obtained by the Planck Collaboration within 1-$\sigma$ errors (see Table \ref{tab:cosmo}).

\begin{figure}
 \includegraphics[width=0.48\textwidth]{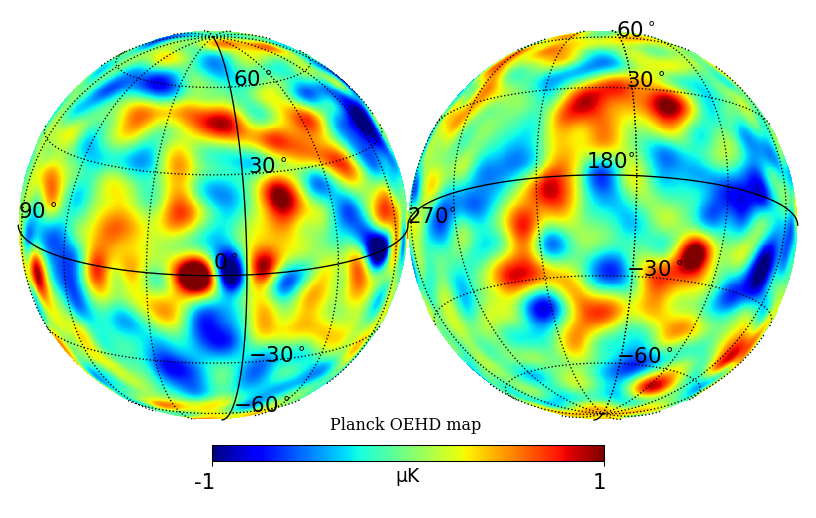}
 \caption{Planck 2018 Odd-Even rings half-mission half-difference  (OEHD) SMICA temperature map, smoothed with 10 degree FWHM Gaussian beam, shown in orthographic projection. This map represents an estimate of the anisotropic noise and residual systematics from the data.}
  \label{fig:noisemap}
\end{figure}

As for the FFP10 anisotropic noise realizations, that incorporate the Planck scanning strategy, Figure \ref{fig:noisecls} shows that the amplitude of the noise for the OE data cuts used (which in turn is in good agreement with a simple noise and residual systematics estimate from the half-mission half-difference OE maps, specially at thigh multipoles), is negligibly small, of order $1\%$ at most, compared to the 1-$\sigma$ errors of the estimated mean power spectrum in the disc-like patches that we use to derive the cosmological parameter maps. In addition, in Figure \ref{fig:noisemap} we show the spatial distribution of the estimated noise from the data (i.e, half-mission half-difference OE maps, smoothed with a 10 degree FWHM Gaussian beam for clarity), is spatially uncorrelated with the "horizons" in the cosmological parameter maps (see e.g, Figures \ref{fig:horizonsh},\ref{fig:horizonsol}). We can thus safely conclude that the anisotropic noise does not have any impact on our main results.

\begin{figure*}
 \includegraphics[width=1.\textwidth]{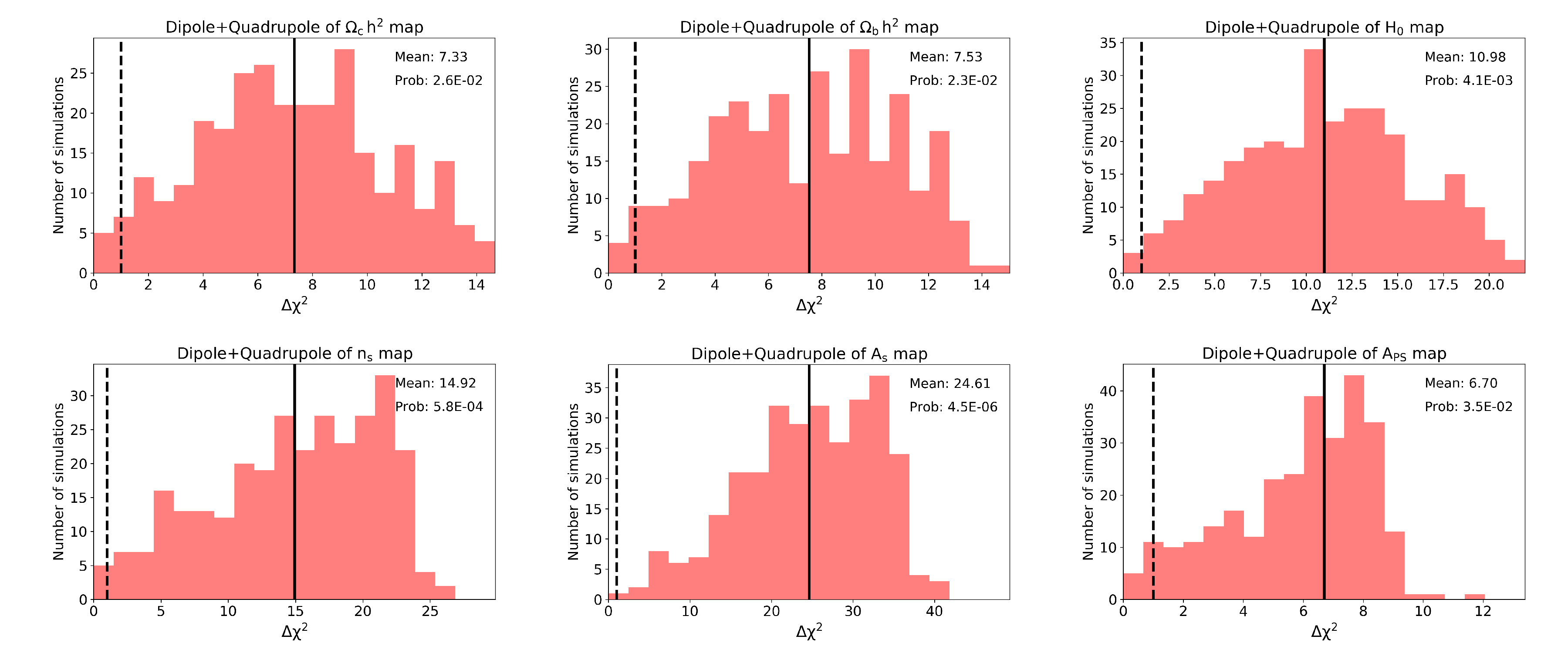}
 \caption{Significance of features in data parameter maps: histogram shows the $\Delta \chi^2$ distribution of the dipole and quadrupole amplitudes in data vs. Gaussian isotropic simulations. Mean values of the distribution in excess from $\Delta \chi^2 =1$ (vertical dashed line) quantify deviations from Gaussian isotropy (dashed line), with a probability of being a statistical fluctuation as quoted in the panel labels.} 
 \label{fig:significance}
\end{figure*}

As an alternative, 
we produced a suite of 300 Gaussian simulations, including the lensed CMB temperature plus a residual foreground sky, making use of the $\tt{SYNFAST}$ HEALPix routine with best-fit parameters to Planck temperature and low multipoles of the polarization data (see Table 2 in \citealt{P18cosmo}; see also Table \ref{tab:cosmo} above), and added one Poisson-like "residual foreground" amplitude that effectively accounts for the combination of CIB and extragalactic point sources contributions from different frequency channels, as explained above (see Section \S\ref{sec:method}). This input foreground amplitude is determined from a joint fit of the basic $\Lambda$CDM cosmology and this "effective nuisance" parameter to the Planck temperature map over the full footprint (i.e, we use the "common" mask for half-mission maps and the "missing pixel" mask for the "Odd-Even" data cuts, which leaves about 76 $\%$ of the sky unmasked). Using this set-up we estimate an amplitude of $\rm{A_{PS}}\simeq 70$, from the SMICA version of the CMB map, as shown in Table \ref{tab:cosmo}.  To the resulting suite of simulated maps, we apply the same methodology for the power spectra estimation in discs across the sky and corresponding cosmological best-fit parameter estimation than we did for the data maps, as described in section \S\ref{sec:method}.

Figure \ref{fig:clsparams} shows the the power spectra of the parameter variation maps from the set of discs is largely dominated by the very lowest multipoles (dipole and quadrupole in particular) and higher multipoles are quickly suppressed i.e, we can use a course sampling of the sky with discs to accurately estimate the observed variations of the parameter maps. In practice, we use discs located at the center of the Healpix pixels of a map with $\rm{Nside=4}$ (i.e, 192 discs across the sky) to measure power spectra of the parameter maps up to $\ell_{max}=12$.  In turn this means that we have estimated a total of $192$ discs/realization $\times$  $300$ realizations $ = 57,600$ 6-dimensional ($\Lambda$CDM+nuisance) parameter fits. Using this procedure, we see that the measured amplitudes for the dipole and quadrupole in the data are clearly in excess of what is found in the Gaussian simulations.

\subsection{Significance of Casual Horizons}
\label{sec:hsignificance}

The appearance of horizons in the angular fluctuations across the sky of the cosmological best-fit values seem quite prominent and their angular size are robust to the mask area used. We note however that the amplitude of the parameter variations depends somewhat on the patch size used, as sample variance modulates them. Ultimately, for the main results of the paper, we have chosen disc size of $60$ degree diameter as the fiducial case which is, as noted in Section \S\ref{sec:horizonsize}, a compromise between sample variance (larger for smaller discs), and over-smoothing (larger for bigger discs).

The significance of the dipole and quadrupole amplitudes of the data parameter map is assessed by estimating how many Gaussian simulations display an amplitude of these two multipoles as large as those measured in the data. In order to quantify this significance, for each parameter $\alpha$, we define a $\Delta \chi^2$ statistic as follows,
\begin{equation}
    \Delta \chi^2_{\alpha} = \sum_{\ell,\ell^{\prime}=1,2} \sum_{i=1}^{N_{sim}} (C^{i}_{\ell, \alpha} - C^{data}_{\ell, \alpha})\cdot Cov^{-1}(\ell,\ell^{\prime}) (C^{i}_{\ell^{\prime}, \alpha} - C^{data}_{\ell^{\prime}, \alpha})
    \label{eq:chi2par}
\end{equation}
where $C^{i}_{\ell}$ and $C^{data}_{\ell}$ with $\ell = 1,2$ refer to the dipole/quadrupole multipoles of the $i$-th simulation and the data, respectively, and $Cov^{-1}(\ell,\ell^{\prime})$ is the inverse of the covariance matrix for the dipole and quadrupole estimated from the set of $N_{sim} = 300$ simulations.

\begin{figure}
 \includegraphics[width=0.48\textwidth]{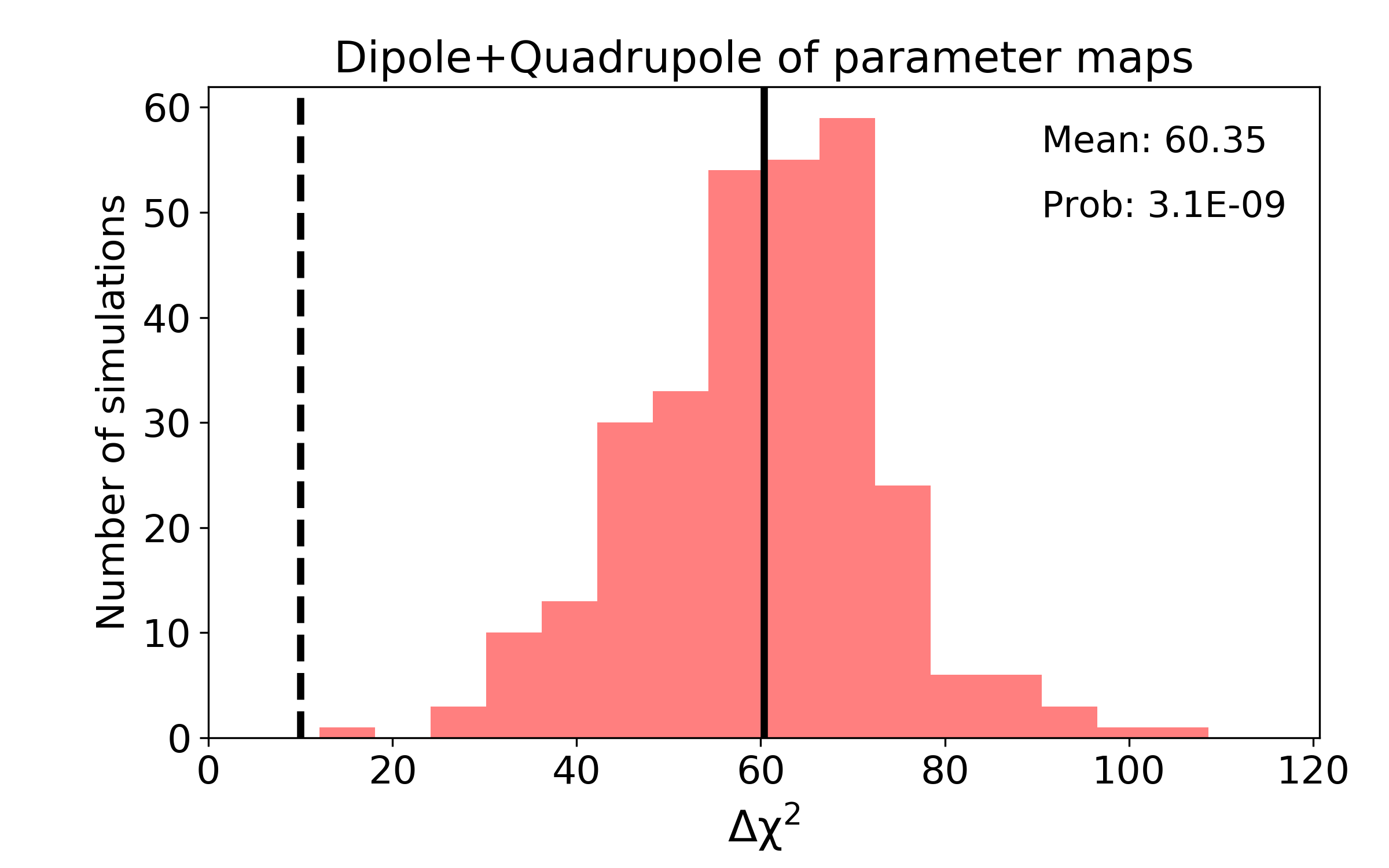}
 \caption{Same as Figure \ref{fig:significance} but for the total significance for the combined $\Lambda$CDM parameters. Solid line displays the significance from the Planck measurement, whereas the dashed line depicts the reference Gaussian isotropic expectation.} 
 \label{fig:chi2tot}
\end{figure}

Our results are summarized in Figure \ref{fig:significance}. By definition, if the data is consistent with a Gaussian realization drawn from this set of simulations, the mean value of this $\Delta \chi^2$ distribution is equal to the number of degrees of freedom, i.e, $2$ (shown as a vertical line in the figure). However the dipole and quadrupole in the data show $\Delta \chi^2$ distributions significantly away from the Gaussian expectations, yielding a $\sim 2.5 \%$ probability of being consistent with a Gaussian fluctuation for 
$\Omega_c h^2$ and $\Omega_b h^2$, whereas this probability drops to $0.4 \%$ for $H_0$, $0.06 \%$ for $n_s$ and $0.0004 \%$ for $A_S$. Although the quoted significances are derived from the SMICA map, we have checked that these values do not significantly change if we use a different component separation map. On the other hand, It is interesting to note that a relatively low probability $3.5 \%$ is also found for the extragalactic residual foreground parameter, $A_{PS}$. However the all-sky mean value (and thus its angular variations) is not robust to the choice of foreground cleaned map used.

We can then estimate a total significance for the cosmological parameter horizons generalizing Eq.(\ref{eq:chi2par}) by combining the 5 cosmological parameter contributions, and including their covariance,
\begin{equation}
    \Delta \chi^2_{Tot} = \sum_{\alpha,\beta} (C^{i}_{\ell, \alpha} - C^{data}_{\ell, \alpha})\cdot Cov^{-1}(\ell,\ell^{\prime},\alpha,\beta) (C^{i}_{\ell^{\prime}, \beta} - C^{data}_{\ell^{\prime}, \beta})
    \label{eq:chi2tot}
\end{equation}
where $\{\alpha,\beta \}$ = $\{\Omega_c\rm{h^2}, \Omega_b\rm{h^2}, \rm{H_0}, \rm{n_s}, \rm{A_s}\}$, and we have omitted the sum over multipoles and simulation index for ease of notation. The result is shown in Figure \ref{fig:chi2tot}. Taking the average over the $\Delta \chi^2_{Tot}$ distribution one thus obtain a value of $60.4$ which for the 10 degrees of freedom  (5 $\Lambda$CDM parameters $\times$ 2 multipoles) gives a Gaussian probability of $3\times 10^{-9}$. It is interesting to note that the covariance between parameters (for a given multipole) is found to have negligible impact (within 2 $\%$) on the computation of the significance, what means that the features of one best-fit cosmological parameter map are not statistically correlated with the corresponding features in a another parameter map, what means that the quoted cosmological anisotropy is signficantly contributed by all the sampled $\Lambda$CDM parameters\footnote{Note that for the covariance estimation we make use of Gaussian isotropic simulations. This makes the amplitude of the power spectrum multipoles of the best-fit cosmological parameter maps from simulations to be systematically lower than the data, as shown in Figure \ref{fig:clsparams}. However for a given high multipole ($\ell > 32$), we expect the {\it Gaussian covariance} between cosmological parameter maps to be a reasonably accurate estimate of the covariance in the data (see \citealt{2018PhRvD..98b3521M} for the  impact of non-Gaussianity, as induced by survey artifacts and foregrounds, in the covariance estimation of CMB "anomalies" at low multipoles).}. On the other hand neglecting the off-diagonal elements of the full covariance, i.e, excluding the covariance between multipoles and among parameters, changes the probability by an order of magnitude (drops from  $3\times 10^{-9}$ to $4\times 10^{-10}$).  This is the main result of this paper, and shows that, {\it for a flat $\Lambda$CDM model, there is a directional dependence of the cosmological parameters across the sky that is inconsistent with the Cosmological principle of isotropy to a very high degree of confidence, i.e, a probability of $\sim 10^{-9}$}. In turn this anisotropy is related to the existence of three well-defined coherent regions, dubbed horizons, where the CMB temperature data prefer significantly different best-fit values.

\subsection{"Tensions" between best-fit parameters from different Horizons}
\label{sec:tensions}

Recent claims in the literature point to significant tensions in cosmological parameter constraints, the most notable being the determination of the Hubble parameter from the CMB data vs. lower redshift probes, such as SN Ia (see \citealt{P18cosmo, Riess19} and references therein). Here we provide an alternative angle to this "tensions" by analyzing the differences one can find already from statistically significant fluctuations across the CMB temperature maps, ie, apparent "tensions" from measurements of the universe at redshift $z\approx 1100$. Because the horizons cover roughly circular patches of $\simeq 60$ degree radius in the CMB sky, the corresponding comoving transverse extent is similar to the comoving radial separation between us and the CMB last scattering surface. Thus if we interpret the measured CMB horizons as causally disconnected regions, we should also expect similar variation in cosmological parameters between $z=0$ and $z \simeq 1100$. 
Figure 4 in \cite{Gazta2021} shows the observable angular size of the horizons as a function of $z$. At $z<2$ this angle is larger than 180 degrees (i.e, the largest accessible separations in the sky), so we do not expect to observe angular variations within the local universe.

\begin{figure*}
 \includegraphics[width=\textwidth]{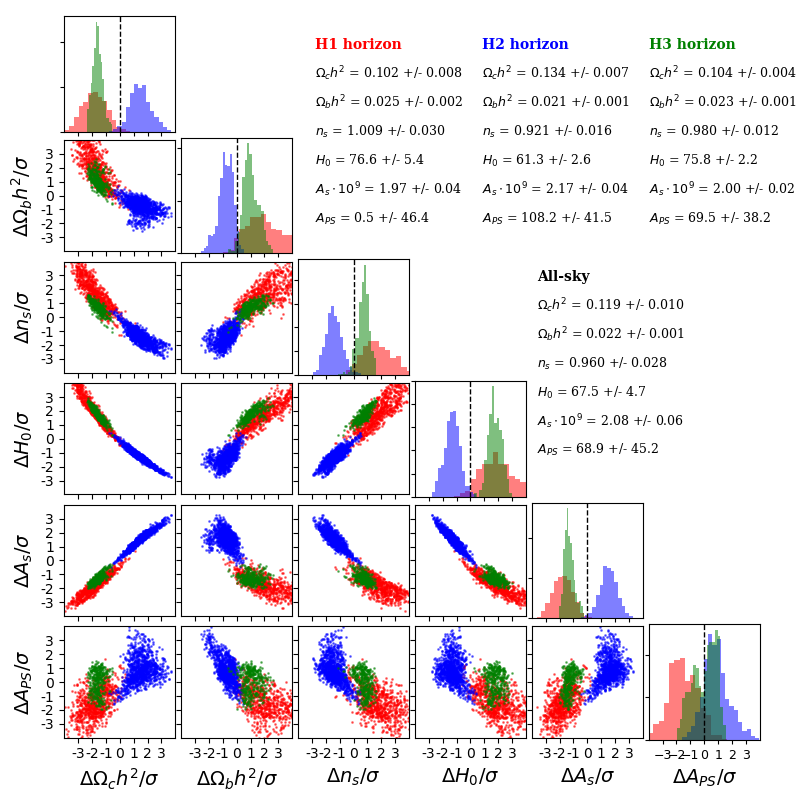}
 \caption{2D parameter correlation plots for measurements within the three horizons identified in the SMICA temperature map, $\rm{H_1}$, $\rm{H_2}$, and $\rm{H_3}$ (scatter plots with colors), and the mean values over the entire footprint (all-sky). For reference, summary statistics include both the mean and scatter for the $12288$ disc measurements across the sky.}
 \label{fig:horizonstats}
\end{figure*}

\begin{figure*}
 \includegraphics[width=\textwidth]{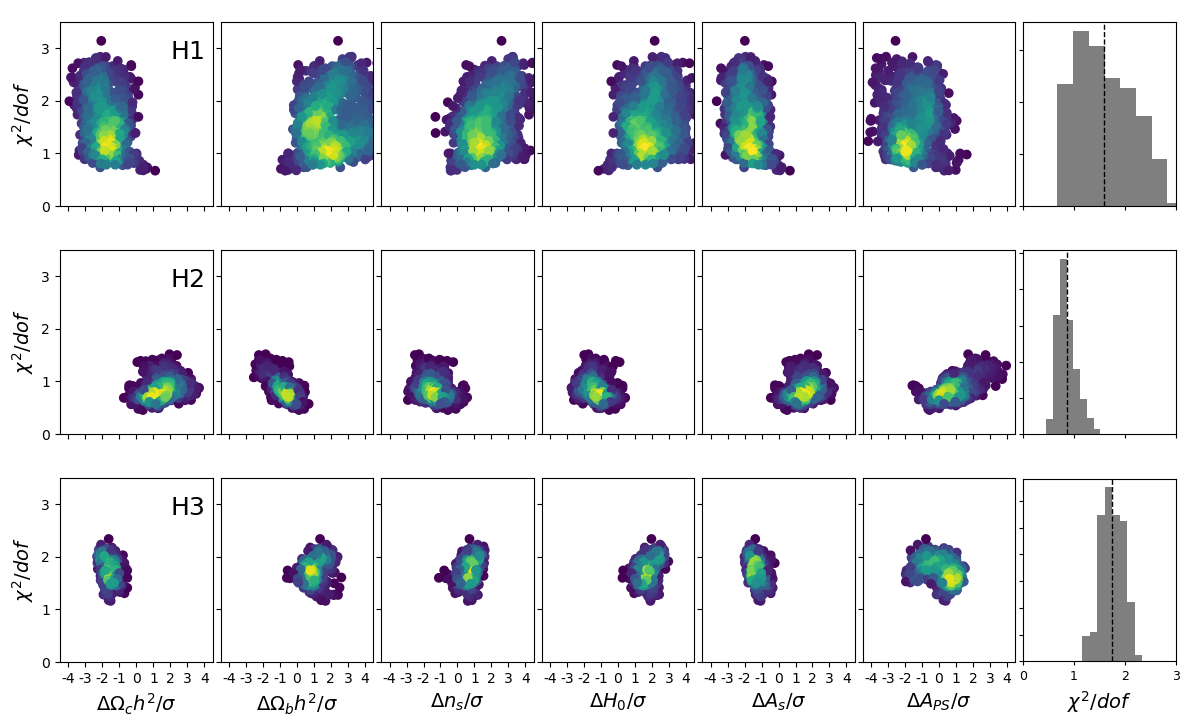}
 \caption{Correlation between the $\chi^2$ per degree of freedom with the best-fit $\Lambda$CDM and residual foreground parameters within the $\rm{H_1}$, $\rm{H_2}$, and $\rm{H_3}$ (from top to bottom rows, respectively). Plots are colored according to density, so that lightest colors correspond to the most highly populated regions in the 2D space (i.e, larger number of disc measurements). The rightmost panels show the distribution of the $\chi^2$ values inside a given horizon, whereas the vertical dashed line displays the mean value.}
 \label{fig:scatter3h}
\end{figure*}

Figure \ref{fig:horizonstats} shows the correlations among cosmological parameters for discs whose pixel at the center is within one well-defined horizon, ie, $\rm{H_1,H_2}$ and $\rm{H_3}$. The first interesting thing to note is that the probability distributions of the parameters hardly overlap for the 2 most significant horizons ($\rm{H_1,H_2}$), with differences significantly away given the errors (rms scatter across patches overlapping each horizon). 

Even if statistically significant, one may wonder whether the horizons we detect are sourced by some unknown systematics instead of a true cosmological signal. In Figure \ref{fig:scatter3h} we display the scatter plots between the $\chi^2/dof$ for each disc measurement across the sky against the $\Lambda$CDM parameters and the nuisance (residual foreground) parameter, for the three horizons detected. Scatter plots are colored according to density, so that lighter colors depict higher density of measurements. Two main features arise from these plots.  First, the fact that most of the discs exhibit low $\chi^2/$dof values (typically $\simeq 1$). This means that the BAO features (which capture most of the information content in the power spectrum in the multipole range considered, $32 < \ell < 2000$) are well fitted by the same flat-space $\Lambda$CDM model that the average measurements over the full Planck footprint, but with slightly different best-fit values. Secondly, if a non-cosmological signal was driving the observed anisotropies, one would expect the largest fluctuations in the cosmological parameter maps would be strongly correlated with high $\chi^2$ values, which is not what we observe. The above two arguments lend further support to the cosmological origin of the observed large-scale parameter anisotropies.

Therefore, if cosmological parameters show significant variations between experiments carried out separately over those particular horizons, one would expect to find statistical "tensions" in the corresponding best-fit parameters. In fact, as pointed out in Section \S\ref{sec:cosmoparams}, parameter differences between the two most significant horizons are up to $\sim 30 \%$, e.g, for the Hubble parameter. If these differences are typical of causally disjoint horizons in the universe, not only at the last scattering surface, as we find here, but also in the local universe, it may provide a simple explanation for the observed cosmological parameter "tensions" between low and high redshift datasets (see e.g, \citealt{Riessetal19,Holicow19,2019arXiv190609189M,P18cosmo,2020MNRAS.tmp.2035P} and references therein).

\section{Discussion}
\label{sec:discussion}

The so-called "anomalies" in CMB data have posed 
a challenge to the standard model since the release of WMAP data, almost two decades ago \citep{WMAP1}. Although the significance of such anomalies depends on the estimator and specific data set used, some of the statistical tensions with the $\Lambda$CDM model have been recently confirmed using the unprecedented quality of the Planck Legacy data \citep{Plegacy}. 

In this paper we address one of the most notable anomalies, the CMB power asymmetry, from a new angle. By analyzing the Planck temperature anisotropy maps in small finite patches, we find that the best-fit cosmological parameters for a simple $\Lambda$CDM model do show evidence for coherent variations across the sky, that we dub "horizons". Although the estimated significance of these horizons varies among the basic set of cosmological parameters, we found that the probability of being consistent with a Gaussian fluctuation is $< 3 \%$ for each of the cosmological parameters explored (see Figure \ref{fig:significance}), which is in qualitative agreement with the quoted significance from previous analysis on CMB anomalies. Moreover, when all parameters are combined, the resulting probability drops to  $\sim 10^{-9}$ (see Figure \ref{fig:chi2tot}), which is a very strong evidence for Cosmological anisotropy on the largest accessible scales.

We shall stress that the anisotropy we find is correlated with the dominant dipole-modulation found in other analyses (such as the power asymmetry analyses), but it is uncorrelated (directionally) with the residual Doppler signal present in the Planck maps. In fact, as shown in \cite{P15isotropy} (see their Figures 34 $\&$ 35), the Doppler residual only affects high multipoles through the so-called "aberration" effect, which points to the direction (l,b) $\sim (260,40)$ degrees, clearly away from our three horizon centers. Moreover, as mentioned in \cite{P18isotropy} (see section 6.2), the asymmetry induced by the Doppler residual modulation has smaller amplitude than the dominant dipole-modulation pattern and thus it has even lower significance. 
Similarly, since our simulations include the lensing signal of the CMB temperature, lensing can not produce the cosmological anisotropy pattern we detect. 

We have tested the robustness of our main results with respect to foreground separation methods. In particular, we found that only the spatial-variation map of the nuisance parameter that encodes residual small-scale foreground contamination changes as a function of the method, unlike the cosmological parameter maps. As a further test, we have also shown that the evidence for horizons in Planck temperature maps is also robust to other possible contributions to the foreground template used, such as dust emission on large scales (see discussion in Section \S\ref{sec:foregrounds}, and Figures \ref{fig:CMmaps}-\ref{fig:parammaps8d}). A quantitative assessment of the impact of foregrounds on the angular power spectrum of the cosmological parameter maps is summarized in Figure \ref{fig:clsparams} showing the robustness of our results to the choice of component separation algorithm.
Moreover, a comparison between Planck and WMAP results (see Section \S\ref{sec:wmap}) showed reasonable good agreement, albeit with some differences, as expected from the lower signal to noise in WMAP. This in turn points to the robustness of the detected anisotropy patterns for experiments with different frequency coverage, analysis pipelines, systematics and foreground separation methods. 

\begin{figure}
 \includegraphics[width=0.48\textwidth]{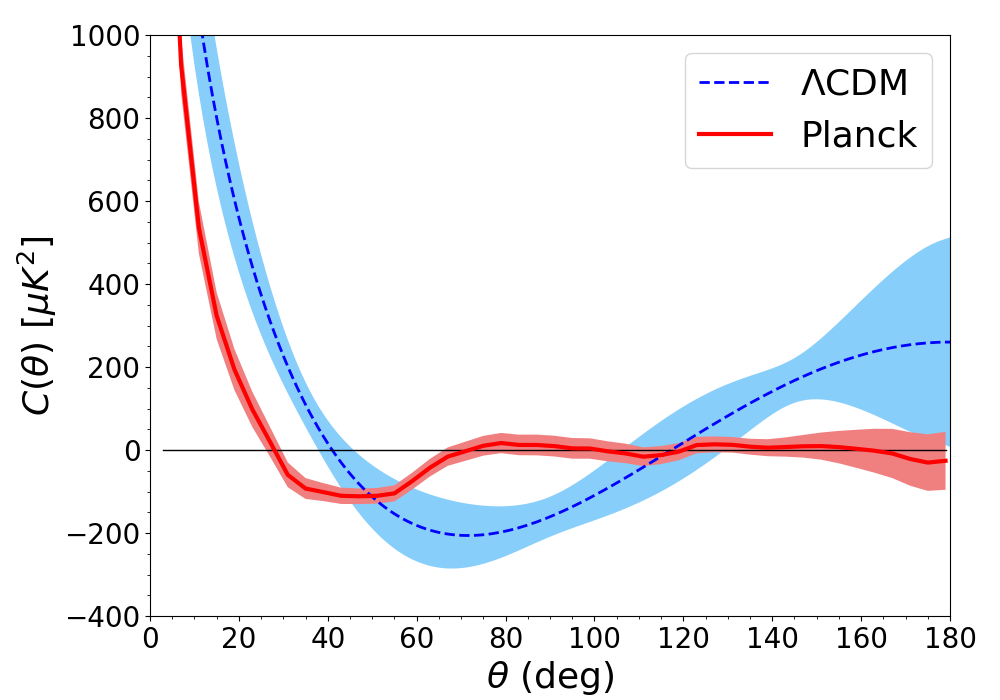}
 \caption{Angular 2-point correlation function of the Planck temperature map (red solid line). For reference we also show the theory prediction for the Planck best-fit $\Lambda$CDM cosmology (blue dashed line). Shaded areas display the 68 $\%$ Gaussian confidence intervals (see text for details).}
 \label{fig:wtheta}
\end{figure}

The estimated size of of these horizons range from $\sim 40$ to $70$ degrees in diameter, i.e, comparable to the largest scales where non-zero correlations are measured from CMB data (see e.g, \citealt{P18cosmo}). In particular, Figure \ref{fig:wtheta} shows the measurement of the angular 2-point correlation function from the Planck SMICA temperature map over the "common" mask that we have used in the main analysis of this paper, which leaves about $76 \%$ of the sky for cosmological analysis. Although angular scales are strongly correlated, our analysis shows that the signal is consistent with zero for angular separations larger than $\sim 65$ degrees, in agreement with previous analyses using WMAP and Planck data \citep{WMAP1,G03,2015MNRAS.451.2978C,2016CQGra..33r4001S}. This lack of large-angle correlations is at variance, given the statistical errors, with the best-fit $\Lambda$CDM model to Planck data\footnote{We estimate the statistical errorbars from the diagonal of the analytic Gaussian covariance matrix (see e.g. Eq.(16) in \citealt{Cabre07}). In particular, these theory errors use as input the $C_{\ell}$'s which are then Legendre transformed to get the corresponding errors in configuration space. Therefore, for consistency, we use the $C_{\ell}$'s measured from the data for the "Planck" case, and the theory ones according to the Planck best-fit cosmology for the "$\Lambda$CDM" case (see Table \ref{tab:cosmo})}.

\begin{figure}
\centering\includegraphics[width=1.\linewidth]{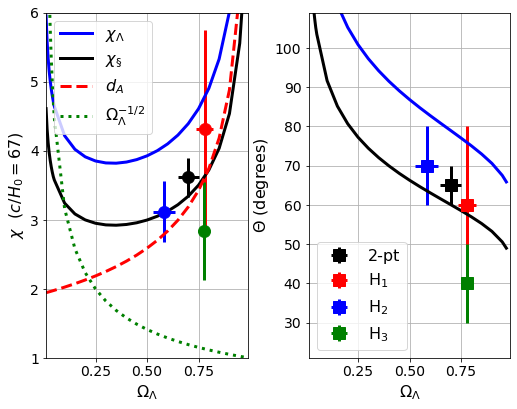}
\caption{
Left panel shows the $\Lambda$ ($\chi_\Lambda$ , blue lines) and causal ( $\chi_\S$, black lines) comoving horizons (in units of $c/H_0$, with $H_0=67$km/s/Mpc) compared to the Hubble horizon $\Omega_\Lambda^{-1/2}$ (dotted line) and the angular diameter distance $d_A$
(dashed red line). As we increase $\Omega_\Lambda$, $\chi$ transits from $\Omega_\Lambda^{-1/2}$ to $d_A$,
which increases with $\Omega_\Lambda$. The right panel shows the corresponding angles $\theta=\chi/d_A$. 
Points with errors show the measurements in the CMB.}
\label{fig:CMBhorizons}
\end{figure}

From our analysis of the Planck temperature map, we find 
$\Omega_\Lambda \simeq 0.8$ on average for $H_1$ and $H_3$, which have estimated sizes of $\theta(H_1) \simeq 60$ deg. and $\theta(H_3) \simeq 40$ deg. respectively, and we find $\Omega_\Lambda \simeq 0.6$ on average for $H_2$, which has an estimated size of $\theta(H_2) \simeq 70$ deg. Right panel of Figure \ref{fig:CMBhorizons}  shows the measurements for $\theta$ (in degrees)
as a function of $\Omega_\Lambda$.
We also estimate the individual comoving scale  $\chi$ for each horizon as $\chi  = \theta d_A$, where $d_A$ is the angular diameter distance estimated with the cosmological parameters in each horizon (see Fig.\ref{fig:horizonstats}). Results are shown in the left panel of Figure \ref{fig:CMBhorizons} in units of $c/H_0$, with $H_0=67$ Km/s/Mpc. Note that the estimates  for $\chi$ are independent from the ones for $\theta$ because the latter are independent of $H_0$.

A $\Lambda$CDM universe is dynamically trapped inside the Hubble horizon $c/H = \Omega_\Lambda^{-1/2} \frac{c}{H_0}$. In comoving coordinates this results in a maximum scale $\chi_\Lambda$ and angle $\theta_\Lambda$:
\begin{equation}
\chi_\Lambda \equiv \int_0^\infty ~\frac{d\ln{a}}{a H(a)} ~~\Rightarrow
~~\theta_\Lambda \equiv \frac{\chi_\Lambda}{d_A} ~~;~~
d_A \equiv \int_{a}^{1} ~\frac{d\ln{a}}{a H(a)} 
\label{eq:sizevsol}
\end{equation}
For a flat $\Omega_\Lambda \simeq 0.7$ we have $\chi_\Lambda \simeq 4.4 c/H_0$. This is close to the observable universe today. 
So in transverse distances, this is a natural scale above which we could expect some anomalies in the CMB sky: $\theta_\Lambda \equiv \chi_\Lambda/d_A  \simeq 79$ degrees, where $d_A$ is the comoving angular diameter distance to the CMB last scattering surface $a \simeq 10^{-3}$. The values of $\chi_\Lambda$ and  $\theta_\Lambda$ (blue lines in Figure \ref{fig:CMBhorizons}) are close, but  above, our horizon measurements.

\cite{G20,Gazta2021,Gazta2021b} has interpreted cosmic acceleration as the result of causally disjoint cosmological horizons or Black Hole Universes (BHU) of size comparable to $\chi_\Lambda$. Such BHU model could simply explain why dark-energy density is comparable to the dark-matter density today (the "why now" issue) and address apparent cosmological parameter tensions (as $H_0$ estimate from low versus high redshift data) as a consequence of the difference in parameter values in disjoint BHU horizons. In particular, as discussed in \cite{G20,Gazta2021}, the causal boundary $\chi_\S$ corresponding to a given $\Omega_\Lambda$ is slightly smaller than the $\Lambda$ trapped surface in Eq.\ref{eq:sizevsol}.This results from the zero action principle, which relates $\Omega_\Lambda$ with the average of matter and radiation within the light-cone.  The relation between $\Omega_\Lambda$ and the size of the BHU is a key prediction of the model that we can test with the measured CMB horizons.

Black lines in Figure \ref{fig:CMBhorizons} show
the prediction for $\chi_\S$ and $\theta_\S$ as a function of $\Omega_\Lambda$ from Eq.31 in \citealt{Gazta2021}.  There is a very good agreement with the CMB measurements.
We also show the "all-sky" measurement of the Planck 2-point angular correlation function (black symbols labeled as "2-pt"). This "all-sky" measurement yields vanishing correlations for scales above $\theta \simeq 65$ degrees for a best-fit value of $\Omega_{\Lambda} \approx 0.7$ (see Table \ref{tab:cosmo}), which is also consistent with the predictions from  \cite{G20,Gazta2021}.
Note that there is no free parameter in these predictions which where published \citep{G20} before the CMB analysis presented here was done.

The good agreement with the BHU predictions, in both angular size and comoving scale,
supports the idea of a physical origin for the anisotropies that we found in the cosmological parameter maps.  The fact that we find a good fit to the same physical model of BAO (but with different parameters in different regions of the sky, see Figure \ref{fig:scatter3h}) makes it hard to explain these anisotropies from systematics or any other non-cosmological signals. It also indicates that the same underlying physical laws apply to different BHU horizons. 

In this paper we have not included polarization data, which could provide another handle on the evidence for anisotropy in cosmological parameter estimation (see e.g \citealt{2010MNRAS.407..399P,2015PhRvD..91f2002M,2016PhRvL.116v1301M}), and constrain the impact of the optical depth to reionization, $\tau$, i.e, possible "patchy" reionization \citep{1996A&A...311....1A,1998ApJ...503..505H,1998ApJ...508..435G,1999ASPC..181..227H} on our results (Fosalba et al. in prep). However we anticipate that since $\tau$ is strongly correlated with the primordial power spectrum amplitude parameter, $\rm{A_S}$, that we have included in our temperature analysis, a priori we do not expect this additional parameter to have a significant impact. In addition, the scale-cuts used ($\ell > 32$) should limit the main expected contribution from the reionization signal.

In summary, we have found strong evidence for a violation of the cosmological principle of isotropy from the analysis of the Planck 2018 temperature map. Our analysis points to significant deviations from statistical isotropy on cosmological scales, with a probability $\sim 10^{-9}$ of being a Gaussian fluctuation. This is the largest reported evidence for a violation of the Cosmological principle to our knowledge.  These parameter variations are consistent with the existence of three distinct patches or horizons with significantly different values with respect to the the mean over the CMB sky.  If the existence of such horizons is confirmed in future analyses (e.g., in high-quality polarization data) this could lend further support to models that predict the existence of those horizons, such as the  \cite{G20,Gazta2021,Gazta2021b} model. This in turn would open the door to unveil the nature of dark-energy and cosmic acceleration, and resolve apparent cosmological parameter tensions reported in recent analyses that combined low and high redshift probes, without the need to invoke new physics beyond our standard model.

\section*{Acknowledgements}
\addcontentsline{toc}{section}{Acknowledgements}
This work is dedicated to the memory of my father, Marcelo.
PF is specially grateful to K.Benabed and O.Dor\'e for useful comments on the draft, I.Tutusaus for advice with $\tt{iMinuit}$, and J.Guerrero for help with the Hidra computing cluster at ICE. Hidra is funded by CSIC project EQC2019-005664-P, with european FEDER funds. The development of this project required $800,000+$ CPU hours at Hidra. Some of the results in this paper have been derived using the $\texttt{healpy}$ \citep{healpy}, $\texttt{HEALPix}$ \citep{Healpix} and $\texttt{CAMB}$ \citep{CAMB} packages.
We acknowledge support from MINECO through grants ESP2017-89838-C3-1-R and PGC2019-102021-B-100, the H2020 European Union grants LACEGAL 734374 and EWC 776247 with ERDF funds, and Generalitat de Catalunya through CERCA to grant 2017-SGR-885 and funding to IEEC.


\section*{Data Availability Statement}

The data underlying this article will be shared on reasonable request to the corresponding author.



\bibliographystyle{mnras}
\bibliography{horizons} 





\bsp	
\label{lastpage}
\end{document}